\documentclass[11pt,a4paper]{article}%
\usepackage{amsfonts}
\usepackage{amsmath}
\usepackage{amssymb}
\usepackage{graphicx}
\usepackage{geometry}%
\providecommand{\U}[1]{\protect\rule{.1in}{.1in}}
\newtheorem{theorem}{Theorem}

\makeatletter
\def \@removefromreset#1#2{\let \@tempb \@elt
\def \@tempa#1{@&#1}\expandafter \let \csname @*#1*\endcsname \@tempa
\def \@elt##1{\expandafter \ifx \csname @*##1*\endcsname \@tempa \else
\noexpand \@elt{##1}\fi}     \expandafter \edef \csname cl@#2\endcsname{\csname cl@#2\endcsname}     \let \@elt \@tempb
\expandafter \let \csname @*#1*\endcsname \@undefined}

\@removefromreset{equation}{section}

\@removefromreset{theorem}{section}
\makeatother
\begin{document}

\title{Specifying the unitary evolution of a qudit for a general nonstationary
Hamiltonian via the generalized Gell-Mann representation}
\author{$^{1,2}$Elena R. Loubenets and $^{1}$Christian K\"{a}ding\\$^{1}$Applied Mathematics Department, National Research University \\Higher School of Economics, Moscow, 101000, Russia, \\$^{2}$Steklov Mathematical Institute of Russian Academy of Sciences, \\Moscow 119991, Russia}
\maketitle

\begin{abstract}
Optimal realizations of quantum technology tasks lead to the necessity of a
detailed analytical study of the behavior of a $d$-level quantum system
(qudit) under a time-dependent Hamiltonian. In the present article, we
introduce a new general formalism describing the unitary evolution of a qudit
$(d\geq2)$ in terms of the Bloch-like vector space and specify how in a
general case this formalism is related to finding time-dependent parameters in
the exponential representation of the evolution operator under an arbitrary
time-dependent Hamiltonian. Applying this new general formalism to a qubit
case $(d=2)$, we specify the unitary evolution of a qubit via the evolution of
a unit vector in $\mathbb{R}^{4}$ and this allows us to derive the precise
analytical expression of the qubit unitary evolution operator for a wide class
of nonstationary Hamiltonians. This new analytical expression includes the
qubit solutions known in the literature only as particular cases.

\end{abstract}

\section{Introduction}

Optimal realizations of many quantum technology tasks need a detailed analysis
of the evolution of a $d\geq2$ dimensional quantum system (a qudit) under a
time-dependent Hamiltonian $H(t)$. In mathematical terms, the evolution of a
qudit under a Hamiltonian $H(t)$ is described on the complex Hilbert space
$\mathbb{C}^{d}$ by the unitary operator $U_{H}(t,t_{0})$ -- the solution of
the Cauchy problem for the nonstationary Schr\"{o}dinger equation with the
initial condition $U_{H}(t_{0},t_{0})=\mathbb{I}$. For a time-independent
Hamiltonian $H$, the solution of this Cauchy problem is well-known and reads
$U_{H}(t,t_{0})=\exp\left\{  -iH(t-t_{0})\right\}  $.

If a Hamiltonian $H(t)$ depends on time, then $U_{H}(t,t_{0})$ is formally
given by the $T$-chronological exponent \cite{Wu, agr} -- the infinite
Volterra series (see Eq. (\ref{3}) in Section 2) -- which however converges
only under some suitable conditions on $H(t).$ For some nonstationary
Hamiltonians beyond these conditions, the analytical expressions for
$U_{H}(t,t_{0})$ via parameters of $H(t)$ are also known, for example, for a
free electron \cite{Flu} in a magnetic field spinning around the
$\mathrm{x}_{3}$-axis. However, for an arbitrary time-dependent $H(t)$, the
analytical expression for $U_{H}(t,t_{0})$ via parameters of $H(t)$ is not
known even in a qubit case.

On the other hand, every unitary operator $V$ on the complex Hilbert space
$\mathbb{C}^{d}$ has the form $\exp\{-i\alpha\}\widetilde{V},$ $\alpha
\in\mathbb{R}$, where a unitary operator $\widetilde{V}$ is an element of the
$SU(d)$ group and, hence, admits the exponential parametrization via the
$SU(d)$ group generators. Therefore, for a $d$-dimensional quantum system, the
exponential representation for $U_{H}(t,t_{0})$ must also exist and there
arises the problem of how to determine time-dependent parameters of this
exponential representation via characteristics of a given qudit Hamiltonian
$H(t).$ To our knowledge, the solution of this problem has not been reported
in the literature even for a qubit case.

In this article, we introduce a new general formalism describing the unitary
evolution of a qudit ($d\geq2$) in terms of the Bloch-like vector space and
specify how in a general case this formalism is related to finding
time-dependent parameters in the exponential representation of $U_{H}%
(t,t_{0})$ under an arbitrary time-dependent Hamiltonian.

Applying this general formalism to a qubit case ($d=2$), we specify the
unitary evolution of a qubit via the evolution of a unit vector in
$\mathbb{R}^{4}$ and find the precise analytical expression of $U_{H}%
(t,t_{0})$ for a wide class of nonstationary qubit Hamiltonians. This new
analytical expression includes the qubit solutions known in the literature
only as particular cases.

The article is organized as follows.

In Section 2, we analyze the known representations for $U_{H}(t,t_{0})$ and
discuss the properties of the generalized Gell-Mann
representation\footnote{Different aspects of the Bloch-like representations
for qudits were considered in \cite{Kim, a, Krys, Kram, Loub,d,b,e}.} for an
arbitrary Hamiltonian and an arbitrary unitary operator on $\mathbb{C}^{d}.$

In Section 3, we derive (Theorem 1) the new general equations specifying the
unitary evolution of a qudit ($d\geq2)$ under a Hamiltonian $H(t)$ in terms of
parameters in the generalized Gell-Mann representation and in the exponential
representation of $U_{H}(t,t_{0}).$

In Sections 4 and 5, we specify (Theorem 2) the forms of these new general
equations in a qubit case $(d=2)$ and derive the novel precise analytical
expression of $U_{H}(t,t_{0})$ for a wide class of qubit Hamiltonians $H(t).$

The main results of the article are summarized in Section 6.

\section{Unitary evolution of a qudit ($d\geq2)$}

Let $H(t):$ $\mathbb{C}^{d}\rightarrow\mathbb{C}^{d},$ $H(t)=H^{\dag}(t),$
$d\geq2,$ be a Hamiltonian of a $d$-level quantum system (qudit). The
evolution of a qudit state
\begin{equation}
\rho(t)=U_{H}(t,t_{0})\rho(t_{0})U_{H}^{\dag}(t,t_{0}),\text{ \ \ }t\geq
t_{0}, \label{1}%
\end{equation}
under a Hamiltonian $H(t)$ is determined by the unitary operator
$U_{H}(t,t_{0})$ -- the solution of the Cauchy problem for the nonstationary
Schr\"{o}dinger equation
\begin{align}
i\frac{d}{dt}U_{H}(t,t_{0})  &  =H(t)U_{H}(t,t_{0}),\text{ \ \ }%
t>t_{0},\label{2}\\
U_{H}(t_{0},t_{0})  &  =\mathbb{I},\nonumber
\end{align}
which satisfies the cocycle property
\begin{equation}
U_{H}(t,t_{0})=U_{H}(t,s)U_{H}(s,t_{0}),\text{ \ \ }s\in\lbrack t,t_{0}],
\label{2.1}%
\end{equation}
and is represented by the chronological operator exponent
\begin{align}
U_{H}(t,t_{0})  &  =\mathrm{T}\exp\left\{  -i\int_{t_{0}}^{t}H(\tau
)d\tau\right\}  :=\label{3}\\
&  =\mathbb{I}-i\int_{t_{0}}^{t}H(\tau)d\tau+\frac{1}{2}(-i)^{2}\int_{t_{0}%
}^{t}d\tau_{1}\int_{t_{0}}^{t}d\tau_{2}\mathrm{T}\left\{  H(\tau_{1}%
)H(\tau_{2})\right\}  +...\nonumber\\
&  +\frac{1}{n!}(-i)^{n}\int_{t_{0}}^{t}d\tau_{1}\int_{t_{0}}^{t}d\tau
_{2}\mathrm{T}\{\text{ }H(\tau_{1})H(\tau_{2})\cdot...\cdot H(\tau
_{n})\}+...,\nonumber
\end{align}
where symbol $\mathrm{T}\{\cdot\}$ means
\begin{equation}
\mathrm{T}\{H(\tau_{1})\cdot...\cdot H(\tau_{m})\}:=H(\tau_{\alpha_{1}}%
)\cdot...\cdot H(\tau_{\alpha_{m}}),\text{ \ \ }\tau_{\alpha_{1}}\geq
\tau_{\alpha_{1}}>...>\tau_{\alpha_{m}}. \label{4}%
\end{equation}
If a Hamiltonian $H(t)$ satisfies the condition
\begin{equation}
\left[  H(t),\int_{t_{0}}^{t}H(\tau)d\tau\right]  =0,\text{ \ }t>t_{0}%
,\text{\ } \label{5}%
\end{equation}
then the series in (\ref{2}) reduces to
\begin{equation}
U_{H}(t,t_{0})=\exp\left\{  -i\int_{t_{0}}^{t}H(\tau)d\tau\right\}  .
\label{6}%
\end{equation}

Recall (see, for example, in \cite{Kim, a, Krys, Kram, Loub}) that any linear
operator $A$ on $\mathbb{C}^{d}$ admits the representation via the generalized
Gell-Mann matrices -- the generalized Gell-Mann representation:%
\begin{align}
A  &  =a_{0}\text{ }\mathbb{I}\text{ \ }\mathbb{+}\text{ \ }\sqrt{\frac{d}{2}%
}\text{ }a\cdot\Lambda,\text{ \ \ \ }a\cdot\Lambda:=\sum_{j=1,...,d^{2}%
-1}a_{j}\cdot\Lambda_{j},\label{7}\\
a_{0}  &  =\frac{1}{d}\mathrm{tr}[A]\in\mathbb{C},\text{ \ \ }a_{j}=\frac
{1}{\sqrt{2d}}\mathrm{tr}[A\Lambda_{j}]\in\mathbb{C},\text{ \ \ }%
a=(a_{1},...a_{d^{2}-1}),\nonumber
\end{align}
where $\Lambda=(\Lambda_{1},...\Lambda_{d^{2}-1})$ is a tuple of traceless
Hermitian operators on $\mathbb{C}^{d}$:%
\begin{equation}
\Lambda_{k}=\Lambda_{k}^{\dag},\text{ \ \ }\mathrm{tr}[\Lambda_{k}%
]=0,\text{\ \ \ }k=1,...,(d^{2}-1), \label{8}%
\end{equation}
satisfying the relations%
\begin{align}
\Lambda_{k}\Lambda_{m}  &  =\frac{2}{d}\delta_{km}\text{ }\mathbb{I+}\sum
_{j}(d_{kmj}+if_{kmj})\Lambda_{j},\label{9}\\
\lbrack\Lambda_{k},\Lambda_{m}]  &  =2i\sum_{j}f_{kmj}\Lambda_{j},\text{
\ \ \ }\mathrm{tr}(\Lambda_{k}\Lambda_{m})=2\delta_{km},\nonumber
\end{align}
and constituting generators of group $\mathrm{SU}(d)$. In (\ref{9}),
$\delta_{km}$ is the Kronecker symbol and $f_{jkm}$, $d_{jkm}$ are
antisymmetric and symmetric structure coefficients of $\mathrm{SU}(d),$
respectively. The matrix representations of the operators $\Lambda_{j},$
$j=1,....,(d^{2}-1),$ in the computational basis of $\mathbb{C}^{d}$
constitute the higher-dimensional extensions of the Pauli matrices in the
qubit case ($d=2$) and the Gell-Mann matrices in the qutrit case ($d=3$).

Under the normalization chosen\footnote{We take the same normalization of a
vector $a$ in representation (\ref{7}) as for traceless qudit observables in
\cite{Loub}.} for a vector $a$ in (\ref{7}),%
\begin{equation}
\mathrm{tr}\left[  A^{\dag}A\right]  =d\left(  \left\vert a_{0}\right\vert
^{2}+\left\Vert a^{\prime}\right\Vert _{\mathbb{C}^{d^{2}-1}}^{2}\right)  .
\label{10}%
\end{equation}
Here and in what follows, by the upper prime $r^{\prime}\in\mathbb{C}%
^{d^{2}-1}$ we denote the column-vector comprised of components of a vector
$r=(r_{1},...,r_{d^{2}-1})$.

Note that representation (\ref{7}) constitutes the decomposition of a linear
operator $A$ on $\mathbb{C}^{d}$ in the orthogonal basis
\begin{equation}
\left\{  \text{ }\mathbb{I},\Lambda_{1},...,\Lambda_{d^{2}-1}\right\}
\label{11}%
\end{equation}
of the vector space $\mathcal{L}$ where linear operators $A:\mathbb{C}%
^{d}\rightarrow\mathbb{C}^{d}$ constitute vectors and the scalar product is
defined by $\langle A_{1},A_{2}\rangle_{\mathcal{L}}:=\mathrm{tr}[A_{1}^{\dag
}A_{2}].$

For a Hamiltonian $H(t)$ on $\mathbb{C}^{d},$ the generalized Gell-Mann
representation (\ref{7}) reads
\begin{align}
H(t)  &  =b_{0}(t)\mathbb{I}\text{ \ }\mathbb{+}\text{ \ }\sqrt{\frac{d}{2}%
}\left(  b_{H}(t)\cdot\Lambda\right)  ,\label{12}\\
b_{0}(t)  &  =\frac{1}{d}\text{\textrm{tr}}[H(t)]\in\mathbb{R},\text{
\ \ }b_{H}^{(j)}(t)=\frac{1}{\sqrt{2d}}\text{\textrm{tr}}[H(t)\Lambda_{j}%
]\in\mathbb{R},\nonumber\\
b_{H}(t)  &  =(b_{H}^{(1)}(t),...,b_{H}^{(d^{2}-1)}(t))\in\mathbb{R}^{d^{2}%
-1},\nonumber
\end{align}
and condition (\ref{5}) implies the following limitations on a vector
$b_{H}(t)\in\mathbb{R}^{d^{2}-1}$:
\begin{equation}%
{\textstyle\sum\limits_{k,m}}
f_{kmj}b_{H}^{(k)}(t)\left(  \int_{t_{0}}^{t}b_{H}^{(m)}(\tau)d\tau\right)
=0,\text{ \ \ }j=1,....,\left(  d^{2}-1\right)  . \label{13}%
\end{equation}
Therefore, if a vector $b_{H}(t)$ satisfies conditions (\ref{13}), then by
(\ref{6})
\begin{equation}
U_{H}(t,t_{0})=\exp\left\{  -i\int_{t_{0}}^{t}b_{0}(\tau)d\tau\right\}
\exp\left\{  -i\sqrt{\frac{d}{2}}\left(  \int_{t_{0}}^{t}b_{H}(\tau
)d\tau\right)  \cdot\Lambda\right\}  . \label{14}%
\end{equation}

However, for an arbitrary qudit Hamiltonian $H(t)$, condition (\ref{5})
(equivalently, conditions (\ref{13})) does not need to be fulfilled, so that
the exponential representation (\ref{14}) of $U_{H}(t,t_{0})$ via the
decomposition coefficients $b_{0}(t),$ $b_{H}(t)$ of a Hamiltonian $H(t)$ by
(\ref{12}) does not, in general, hold.

On the other hand, as it is the case for every unitary operator on
$\mathbb{C}^{d},$ operator $U_{H}(t,t_{0})$ must have the form%
\begin{equation}
U_{H}(t,t_{0})=\exp\left\{  -i\alpha(t,t_{0})\right\}  \widetilde{U}%
_{H}(t,t_{0}),\text{ \ \ }\alpha(t,t_{0})\in\mathbb{R}, \label{15}%
\end{equation}
where $\widetilde{U}_{H}(t,t_{0})\in\mathrm{SU}(d)$ and, hence, as any element
of $\mathrm{SU}(d)$, admits (see, for example, \cite{Kuzn} and references
therein) the exponential parametrization
\begin{equation}
\widetilde{U}_{H}(t,t_{0})=\exp\left\{  -i\text{ }\sqrt{\frac{d}{2}}\left(
n_{H}(t,t_{0})\cdot\Lambda\right)  \right\}  \label{16}%
\end{equation}
via generators $\Lambda_{1},...\Lambda_{d^{2}-1}$ of group \textrm{SU}%
$(d)\ $and a vector $n_{H}(t,t_{0})=(n_{1},...n_{d^{2}-1})\in\mathbb{R}%
^{d^{2}-1},$ which in case of solution $\widetilde{U}_{H}(t,t_{0}%
)\in\mathrm{SU}(d)$ depends also on a Hamiltonian $H(t),$ time $t$ and an
initial moment $t_{0}$. In (\ref{16}), similarly as in decomposition
(\ref{7}), we use the following normalization for a vector $n_{H}(t,t_{0}):$
\begin{equation}
\mathrm{tr}\left[  \left(  \sqrt{\frac{d}{2}}n_{H}(t,t_{0})\cdot
\Lambda\right)  ^{2}\right]  =d\left\Vert n_{H}(t,t_{0})\right\Vert
_{\mathbb{R}^{d^{2}-1}}^{2}. \label{17}%
\end{equation}

Relations (\ref{15}), (\ref{16}) imply that, for every\footnote{For which a
unique solution of (\ref{2}) exists.} qudit Hamiltonian $H(t)$, the unitary
evolution operator $U_{H}(t,t_{0})$ admits the exponential representation
\begin{align}
U_{H}(t,t_{0})  &  =\exp\left\{  -i\alpha(t,t_{0})\right\}  \exp\left\{
-i\text{ }\sqrt{\frac{d}{2}}\left(  n_{H}(t,t_{0})\cdot\Lambda\right)
\right\}  ,\label{18}\\
\alpha(t_{0},t_{0})  &  =0,\text{ \ \ }n_{H}(t_{0},t_{0})=0,\nonumber
\end{align}
where parameters $\alpha(t,t_{0}),$ $n_{H}(t,t_{0})$ can be presented in the
form
\begin{equation}
\alpha(t,t_{0})=\int_{t_{0}}^{t}\beta_{0}(\tau)d\tau,\text{ \ \ }\beta
_{0}(\tau)\in\mathbb{R},\text{ \ \ }n_{H}(t,t_{0})=\int_{t_{0}}^{t}\beta
_{H}(\tau)d\tau,\text{ \ \ }\beta_{H}(t)\in\mathbb{R}^{d^{2}-1}. \label{19}%
\end{equation}
This implies
\begin{equation}
U_{H}(t,t_{0})=\exp\left\{  -i\int_{t_{0}}^{t}\beta_{0}(\tau)d\tau\right\}
\exp\left\{  -i\text{ }\sqrt{\frac{d}{2}}\left(  \int_{t_{0}}^{t}\beta
_{H}(\tau)d\tau\right)  \text{\ }\cdot\Lambda\right\}  . \label{20}%
\end{equation}
The form of this representation is quite similar to the one of representation
(\ref{14}), which is valid if a Hamiltonian $H(t)$ satisfies condition
(\ref{13}). However, for an arbitrary Hamiltonian $H(t),$ a vector $\beta
_{H}(t)\in\mathbb{R}^{d^{2}-1}$ in (\ref{20}) does not need to be equal to a
vector $b_{H}(t)\in\mathbb{R}^{d^{2}-1}$ in representation (\ref{12}) for this
$H(t).$

Therefore, in order to specify the unitary evolution operator $U_{H}(t,t_{0})$
under an arbitrary nonstationary Hamiltonian $H(t),$ we need to express
parameters $\beta_{0}(t)\in\mathbb{R},$ $\beta_{H}(t)\in\mathbb{R}^{d^{2}-1}$
in (\ref{20}) via coefficients $b_{0}(t)\in\mathbb{R},$ $b_{H}(t)\in
\mathbb{R}^{d^{2}-1}$ in the generalized Gell-Mann representation (\ref{12})
for a given $H(t)$.

In the proceeding sections, we consider this problem for an arbitrary $d\geq2$
and further study the case $d=2$ in detail.

\section{Evolution equations in the Bloch-like vector space}

Together with the generalized Gell-Mann representation (\ref{12}) for a
Hamiltonian $H(t)$, let us also specify decomposition (\ref{7}) for a unitary
operator (\ref{16}) on $\mathbb{C}^{d}$:%
\begin{align}
\widetilde{U}_{H}(t,t_{0})  &  =\exp\left\{  -i\text{ }\sqrt{\frac{d}{2}%
}\left(  n_{H}(t,t_{0})\cdot\Lambda\right)  \right\} \label{21}\\
&  =u_{0}(t,t_{0})\mathbb{I}+\sqrt{\frac{d}{2}}u_{H}(t,t_{0})\cdot
\Lambda,\text{ \ \ }u_{H}(t,t_{0})=(u_{H}^{(1)},...,u_{H}^{(d^{2}%
-1)}),\nonumber\\
u_{0}(t,t_{0})  &  =\frac{1}{d}\mathrm{tr}[\widetilde{U}_{H}(t,t_{0}%
)]\in\mathbb{C},\text{ \ \ }u_{H}^{(j)}(t,t_{0})=\frac{1}{\sqrt{2d}%
}\mathrm{tr}[\widetilde{U}_{H}(t,t_{0})\Lambda_{j}]\in\mathbb{C}.\nonumber
\end{align}
The initial conditions in (\ref{18}) and the unitary property of
$\widetilde{U}_{H}(t,t_{0})$ imply%
\begin{equation}
u_{0}(t_{0},t_{0})=1,\text{ \ \ }u_{H}^{(j)}(t_{0},t_{0})=0, \label{22}%
\end{equation}
and%
\begin{align}
\left\vert u_{0}(t,t_{0})\right\vert ^{2}+\left\Vert u_{H}^{^{\prime}}%
(t,t_{o})\right\Vert _{\mathbb{C}^{d^{2}-1}}^{2}  &  =1,\label{23}\\
0  &  =u_{0}(t,t_{0})\left(  u_{_{H}}^{(j)}(t,t_{0})\right)  ^{\ast}%
+u_{0}^{\ast}(t,t_{0})u_{H}^{(j)}(t,t_{0})\nonumber\\
&  +\sqrt{\frac{d}{2}}\sum_{k,m}\left(  d_{kmj}+if_{kmj}\right)  u_{H}%
^{(k)}(t,t_{0})\left(  u_{H}^{(m)}(t,t_{0})\right)  ^{\ast},\nonumber
\end{align}
for all $t\geq t_{0}$ and all $j=1,...,(d^{2}-1).$

Substituting (\ref{21}) into (\ref{18}) and (\ref{18}) into (\ref{2}) and
taking $u_{H}(t,t_{0})=i\widetilde{u}_{H}(t,t_{0})$, we derive
\begin{equation}
\alpha(t,t_{0})=\exp\left\{  -i\int_{t_{0}}^{t}b_{0}(\tau)d\tau\right\}
\label{24}%
\end{equation}
and the following system of ordinary differential equations for $u_{0}%
(t,t_{0})$ and $\widetilde{u}_{H}(t,t_{0}):$
\begin{align}
\overset{\cdot}{u}_{0}(t,t_{0})  &  =b_{_{H}}(t)\cdot\widetilde{u}_{H}%
(t,t_{0}),\label{25}\\
\frac{d}{dt}\widetilde{u}_{H}^{(j)}(t,t_{0})  &  =-u_{0}(t,t_{0})b_{H}%
^{(j)}+\sqrt{\frac{d}{2}}\sum_{m,k}\left(  \text{ }f_{kmj}-id_{kmj}\right)
b_{H}^{(k)}(t)\widetilde{u}_{H}^{(m)}(t,t_{0}),\nonumber\\
u_{0}(t_{0},t_{0})  &  =1,\text{ \ \ }\widetilde{u}_{H}(t_{0},t_{0}%
)=0.\nonumber
\end{align}
Relations (\ref{23}) constitute the functionally independent first integrals
of these ODEs.

Thus, for an arbitrary $d\geq2,$ the unitary evolution operator $U_{H}%
(t,t_{0})$ under a Hamiltonian $H(t)$ is given by
\begin{align}
U_{H}(t,t_{0})  &  =\exp\left\{  -i\int_{t_{0}}^{t}b_{0}(\tau)d\tau\right\}
\exp\left\{  -i\text{ }\sqrt{\frac{d}{2}}\left(  n_{H}(t,t_{0})\cdot
\Lambda\right)  \right\} \label{26}\\
&  =\exp\left\{  -i\int_{t_{0}}^{t}b_{0}(\tau)d\tau\right\}  \left(
u_{0}(t,t_{0})\text{ }\mathbb{I}\text{\ }\mathbb{+}\text{ }i\text{ }%
\sqrt{\frac{d}{2}}\widetilde{u}_{H}(t,t_{0})\cdot\Lambda\right)  ,\nonumber
\end{align}
where $u_{0}(t,t_{0})\in\mathbb{C},$ $\widetilde{u}_{H}^{\prime}(t,t_{0}%
)\in\mathbb{C}^{d^{2}-1}$ satisfy the Cauchy problem (\ref{25}) for the
nonautonomous system of linear ordinary differential equations (ODEs).

On the other hand, due to the results in \cite{Kuzn}, we can explicitly
represent $u_{0}(t)\in\mathbb{C},$ $\widetilde{u}_{H}^{\prime}(t)\in
\mathbb{C}^{d^{2}-1}$ in (\ref{26}) via a vector $n_{H}(t,t_{0})$.

Namely, for each group element $\mathrm{V}_{d}\in\mathrm{SU}(d)$ with the
exponential parametrization%
\begin{equation}
\mathrm{V}_{d}(r)=\exp\left\{  -i\sqrt{\frac{d}{2}}(r\cdot\Lambda)\right\}
,\text{ \ \ }r\in\mathbb{R}^{d^{2}-1}, \label{27}%
\end{equation}
let us consider the generalized Gell-Mann representation (\ref{7}):
\begin{align}
\exp\left\{  -i\sqrt{\frac{d}{2}}(r\cdot\Lambda)\right\}   &  =\mathrm{v}%
_{0}(r)\text{ }\mathbb{I}\text{ \ }\mathbb{+}\text{ \ }\sqrt{\frac{d}{2}%
}\left(  \mathrm{v}(r)\cdot\Lambda\right)  ,\label{28}\\
\left\vert \mathrm{v}_{0}(t)\right\vert ^{2}+\left\Vert \mathrm{v}^{\prime
}(t)\right\Vert _{\mathbb{C}^{d^{2}-1}}^{2}  &  =1,\nonumber
\end{align}
where
\begin{align}
\mathrm{v}_{0}(r)  &  =\frac{1}{d}\mathrm{tr}\left[  \exp\left\{
-i\sqrt{\frac{d}{2}}(r\cdot\Lambda)\right\}  \right]  ,\label{29}\\
\mathrm{v}(r)  &  =\frac{1}{\sqrt{2d}}\mathrm{tr}\left[  \Lambda\exp\left\{
-i\sqrt{\frac{d}{2}}(r\cdot\Lambda)\right\}  \right]  .\nonumber
\end{align}
Denote by $\mathrm{E}(\lambda_{m}(r))$ the spectral projection of a Hermitian
operator $(r\cdot\Lambda)$ corresponding to its eigenvalue $\lambda_{m}%
(r)\in\mathbb{R}$ with multiplicity $k_{\lambda_{_{m}}(r)}$. The spectral
decomposition of $\mathrm{V}_{d}(r)=\exp\{-i\sqrt{\frac{d}{2}}(r\cdot
\Lambda)\}$ reads%
\begin{equation}
\mathrm{V}_{d}(r)=\sum_{\lambda_{m}}\exp\left\{  -i\sqrt{\frac{d}{2}}%
\lambda_{m}(r)\right\}  \mathrm{E(}\lambda_{m}(r)). \label{30}%
\end{equation}
Substituting this into relations (\ref{29}) and taking into account the cyclic
property of the trace and relation $\mathrm{tr}[\mathrm{E}(\lambda
_{m}(r))]=k_{\lambda_{_{m}}(r)}$, we derive\footnote{These expressions differ
by normalizations from those in \cite{Kuzn}.}
\begin{align}
\mathrm{v}_{0}(r)  &  =\frac{1}{d}\mathrm{tr}\left[  V_{d}(r)\right]
=\frac{1}{d}\sum_{\lambda_{m}}k_{\lambda_{_{m}}(r)}\exp\left\{  -i\sqrt
{\frac{d}{2}}\lambda_{m}(r)\right\}  ,\label{31}\\
& \nonumber\\
\mathrm{v}_{j}(r)  &  =\frac{1}{\sqrt{2d}}\mathrm{tr}\left[  \Lambda_{j}%
V_{d}(r)\right]  =\frac{1}{\sqrt{2d}}\mathrm{tr}\left[  \Lambda_{j}\left(
\sum_{m=0,1,...}\frac{\left(  -i\right)  ^{m}}{m!}\left(  \frac{d}{2}\right)
^{\frac{m}{2}}(r\cdot\Lambda)^{m}\right)  \right] \label{32}\\
&  =\frac{i}{d}\frac{\partial}{\partial r_{j}}\mathrm{tr}\left[
V_{d}(r)\right]  ,\nonumber
\end{align}
which imply%
\begin{align}
\mathrm{v}_{0}(r)  &  =\frac{1}{d}K_{d}(r),\text{ \ \ }\mathrm{v}_{j}%
(r)=\frac{i}{d}\left(  \nabla_{r}\text{ }K_{d}(r)\cdot\Lambda\right)
,\label{33}\\
\mathrm{V}_{d}(r)  &  =\frac{1}{d}K_{d}(r)\mathbb{I+}i\sqrt{\frac{d}{2}%
}\left(  \frac{1}{d}\nabla_{r}\text{ }K_{d}(r)\cdot\Lambda\right)  ,\nonumber
\end{align}
where $\nabla_{r}:=\left(  \frac{\partial}{\partial r_{1}},...,\frac{\partial
}{\partial r_{d^{2}-1}}\right)  $ and$\ $%
\begin{equation}
K_{d}(r):=\sum_{\lambda_{m}(r)}k_{\lambda_{m}(r)}\exp\left\{  -i\sqrt{\frac
{d}{2}}\lambda_{m}(r)\right\}  . \label{34}%
\end{equation}

From (\ref{28}) and (\ref{33}) it follows that, in relation (\ref{26}),
\begin{equation}
u_{0}(t,t_{0})=\frac{1}{d}K_{d}\left(  n_{H}(t,t_{0})\right)  ,\text{
\ \ }\widetilde{u}_{_{H}}(t)=\frac{1}{d}\nabla_{n_{_{H}}}\left(  K_{d}%
(n_{H}(t,t_{0})\right)  ), \label{35}%
\end{equation}
for some vector $n_{H}(t,t_{0})\in\mathbb{R}^{d^{2}-1},$ so that
\begin{align}
\widetilde{U}_{H}(t,t_{0})  &  =\exp\left\{  -i\text{ }\sqrt{\frac{d}{2}%
}\left(  n_{H}(t,t_{0})\cdot\Lambda\right)  \right\} \label{36}\\
&  =\frac{1}{d}K_{d}\left(  n_{H}(t,t_{0})\right)  \text{ }\mathbb{I}\text{
}+\text{ }i\sqrt{\frac{d}{2}}\frac{1}{d}\nabla_{n_{_{H}}}\left(  K_{d}%
(n_{H}(t,t_{0})\right)  )\cdot\Lambda.\nonumber
\end{align}
The substitution of (\ref{35}) into the first and the second equations of the
system of linear ODEs (\ref{25}) gives%
\begin{align}
&  \frac{\partial}{\partial n_{H}}K_{d}(n_{H}(t,t_{0}))\cdot\frac
{dn_{H}(t,t_{0})}{dt}\label{37}\\
&  =b_{H}(t)\cdot\frac{\partial}{\partial n_{H}}K_{d}(n_{H}(t,t_{0}%
))\text{\ }\nonumber\\
&  \Leftrightarrow\nonumber\\
&  \frac{dn_{H}(t,t_{0})}{dt}-b_{H}(t)\text{ \ }\bot\text{ \ }\frac{\partial
}{\partial n_{H}}K_{d}(n_{H}(t,t_{0}))\nonumber
\end{align}
and
\begin{align}
&  \frac{\partial}{\partial n_{H}}\left(  \frac{\partial}{\partial n_{H}%
^{(j)}}\left(  K_{d}(n_{H}(t,t_{0})\right)  \right)  \cdot\frac{dn_{H}%
(t,t_{0})}{dt}\label{38}\\
&  =-b_{H}^{(j)}(t)K_{d}(n_{H}(t,t_{0}))\text{ }+\text{ }\sqrt{\frac{d}{2}%
}\sum_{k,m}\left(  \text{ }f_{kmj}-id_{kmj}\right)  b_{H}^{(k)}(t)\frac
{\partial}{\partial n_{H}^{(m)}}K_{d}(n_{H}(t,t_{0})),\nonumber\\
j  &  =1,...,(d^{2}-1),\nonumber
\end{align}
respectively.

Relations (\ref{18}) and (\ref{21})--(\ref{38}) prove the following statement.

\begin{theorem}
Let $H(t)=b_{0}(t)\mathbb{I}+b(t)\cdot\Lambda,$ $b(t)\in\mathbb{R}^{d^{2}-1},$
be a Hamiltonian on $\mathbb{C}^{d}.$ For each $d\geq2,$ the unitary operator
$U_{H}(t,t_{0})$ on $\mathbb{C}^{d}$ describing the evolution of a qudit under
a Hamiltonian $H(t)$ has the form
\begin{align}
U_{H}(t,t_{0})  &  =\exp\left\{  -i\int_{t_{0}}^{t}b_{0}(\tau)d\tau\right\}
\exp\left\{  -i\text{ }\sqrt{\frac{d}{2}}\left(  n_{H}(t,t_{0})\cdot
\Lambda\right)  \right\} \label{39}\\
&  =\exp\left\{  -i\int_{t_{0}}^{t}b_{0}(\tau)d\tau\right\}  \left(
u_{0}(t,t_{0})\text{ }\mathbb{I}\text{ }+\text{\ }i\sqrt{\frac{d}{2}%
}\widetilde{u}_{_{H}}(t,t_{0})\cdot\Lambda\right)  .\nonumber
\end{align}
Here, the scalar function $u_{0}(t,t_{0})\in\mathbb{C}$ and vector
$u_{H}(t,t_{0})=(u_{H}^{(1)},...,u_{H}^{(d^{2}-1)}),$ $u_{H}^{(j)}%
\in\mathbb{C},$ are the solutions of the Cauchy problem (\ref{25}),
equivalently,
\begin{equation}
u_{0}(t,t_{0})=\frac{1}{d}K_{d}\left(  n_{H}(t,t_{0})\right)
,\ \ \ \widetilde{u}_{H}(t,t_{0})=\frac{1}{d}\nabla_{n_{_{H}}}\left(
K_{d}(n_{H}(t,t_{0})\right)  , \label{40}%
\end{equation}
where function $K_{d}(n)$ is given by (\ref{34}) and vector $n_{H}%
(t)\in\mathbb{R}^{d^{2}-1}$ is the solution of the Cauchy problem
\begin{equation}
\frac{dn_{H}(t,t_{0})}{dt}=b_{H}(t)+n_{\bot}(t,t_{0}),\ \ \ n_{H}(t_{0}%
,t_{0})=0, \label{41}%
\end{equation}
with $n_{\bot}(t,t_{0})\in\mathbb{R}^{d^{2}-1}$ satisfying for all $t>t_{0}$
the orthogonality relation $n_{\bot}(t,t_{0})\cdot\nabla_{n_{H}}(K_{d}%
(n_{H}(t,t_{0}))=0$ and determined via the equation%
\begin{align}
&  \frac{dn_{H}(t,t_{0})}{dt}\cdot\frac{\partial}{\partial n_{H}}\left(
\frac{\partial}{\partial n_{j}}K_{d}(n_{H}(t,t_{0})\right) \label{42}\\
&  =-b_{H}^{(j)}(t)K_{d}(n_{H}(t,t_{0}))\text{ }+\text{ }\sqrt{\frac{d}{2}%
}\sum_{k,m}\left(  \text{ }f_{kmj}-id_{kmj}\right)  b_{H}^{(k)}(t)\frac
{\partial}{\partial n_{H}^{(m)}}K_{d}(n_{H}(t,t_{0})).\nonumber
\end{align}

\end{theorem}

In Sections 4 and 5, we specify Eqs. (\ref{25}), (\ref{41}), (\ref{42}) for a
general qubit case.

\subsection{Finding $K_{d}(n)$ for d=2,3}

In this subsection, we consider the characteristic function $K_{d}(r)$, given
by (\ref{34}), and also, representation (\ref{28}) for $d=2,3.$

\begin{itemize}
\item For $d=2$, the matrix representations of generators $\sigma_{1}%
,\sigma_{2},\sigma_{3}$ of \textrm{SU}$(2)$ in the computational basis in
$\mathbb{C}^{2}$ are given by the Pauli matrices%
\begin{align}
\sigma_{1}  &  =%
\begin{pmatrix}
0 & 1\\
1 & 0
\end{pmatrix}
,\text{ \ \ \ }\sigma_{2}=%
\begin{pmatrix}
0 & -i\\
i & 0
\end{pmatrix}
,\text{ \ \ }\sigma_{3}=%
\begin{pmatrix}
1 & 0\\
0 & -1
\end{pmatrix}
,\label{43}\\
\sigma &  =(\sigma_{1},\sigma_{2},\sigma_{3}),\text{ \ \ \textrm{tr}}%
[\sigma_{k}\sigma_{j}]=2\delta_{jk},\nonumber\\
\sigma_{1}\sigma_{2}  &  =i\sigma_{3},\text{ \ \ }\sigma_{2}\sigma_{3}%
=i\sigma_{1},\text{ \ \ }\sigma_{3}\sigma_{1}=i\sigma_{2},\nonumber
\end{align}
and, for each vector $r\in\mathbb{R}^{3},$ the traceless Hermitian operator
$n\cdot\sigma$ on $\mathbb{C}^{2}$ has eigenvalues $\pm\left\Vert n\right\Vert
_{\mathbb{R}^{3}}.$ Therefore, by (\ref{34}), the characteristic function
$K_{2}(r)$ and its derivatives are given by
\begin{align}
K_{2}(r)  &  =\exp\left\{  -i\text{ }\left\Vert r\right\Vert _{\mathbb{R}^{3}%
}\right\}  +\exp\left\{  i\text{ }\left\Vert r\right\Vert _{\mathbb{R}^{3}%
}\right\} \label{44}\\
&  =2\cos\left\Vert r\right\Vert _{\mathbb{R}^{3}},\nonumber\\
\frac{\partial}{\partial r_{j}}K_{2}(r)  &  =-2\sin\left(  \left\Vert
r\right\Vert _{\mathbb{R}^{3}}\right)  \frac{r_{j}}{\left\Vert r\right\Vert
_{\mathbb{R}^{3}}},\nonumber
\end{align}
and representation (\ref{28}) reduces to the well-known formula%
\begin{equation}
\exp\{-i\text{ }(r\cdot\sigma)\}=\mathbb{I}\cos\left\Vert r\right\Vert
_{\mathbb{R}^{3}}-i\sin\left(  \left\Vert r\right\Vert _{\mathbb{R}^{3}%
}\right)  \frac{r\cdot\sigma}{\left\Vert r\right\Vert _{\mathbb{R}^{3}}},
\label{45}%
\end{equation}
see, for example, in \cite{Nil}.

\item For $d=3,$ the matrix representations of the \textrm{SU}$(3)$ generators
in the computational basis in $\mathbb{C}^{3}$ constitute the Gell-Mann
matrices. For each $r\in\mathbb{R}^{8},$ the traceless Hermitian operator
$\left(  r\cdot\Lambda\right)  $ on $\mathbb{C}^{3}$ has eigenvalues
\cite{Kuzn}%
\begin{align}
\lambda_{1,2}(r)  &  =\frac{2}{\sqrt{3}}\left\Vert r\right\Vert _{\mathbb{R}%
^{8}}\sin\left(  \phi(r)\pm\frac{\pi}{3}\right)  ,\label{46}\\
\lambda_{3}(r)  &  =-\frac{2}{\sqrt{3}}\left\Vert r\right\Vert _{\mathbb{R}%
^{8}}\sin\left(  \phi(r)\right)  ,\nonumber
\end{align}
where
\begin{equation}
\sin\left(  3\phi(r)\right)  =-\frac{3\sqrt{3}}{2\left\Vert r\right\Vert
_{\mathbb{R}^{8}}^{3}}\det(r\cdot\Lambda). \label{46.1}%
\end{equation}

From relations (\ref{34}) and (\ref{46}) it follows that, for $d=3$,
\begin{align}
K_{3}(r)  &  =\exp\left\{  -i\sqrt{\frac{3}{2}}\lambda_{1}(r)\right\}
+\exp\left\{  -i\sqrt{\frac{3}{2}}\text{ }\lambda_{2}(r)\right\}
+\exp\left\{  -i\text{ }\sqrt{\frac{3}{2}}\lambda_{3}(r)\right\} \label{47}\\
&  =\sum\limits_{k=0,1,2}\exp\{-i\sqrt{2}\left\Vert r\right\Vert
_{\mathbb{R}^{8}}\sin(\phi(r)+2\pi k/3)\}\nonumber
\end{align}
and (see Appendix B)
\begin{equation}
\frac{\partial}{\partial r}K_{3}(r)=-3i\sqrt{\frac{2}{3}}\left(
F_{1}(r)\text{ }p(r)+F_{2}(r)\frac{\text{ }r}{\left\Vert r\right\Vert
_{\mathbb{R}^{8}}}\right)  , \label{47.1}%
\end{equation}
where%
\begin{align}
p^{(m)}(r)  &  :=\sum\limits_{i,j=1}^{8}\frac{r^{(i)}r^{(j)}d_{ijm}%
}{\left\Vert r\right\Vert _{\mathbb{R}^{8}}^{2}},\text{ \ \ }p^{\prime}%
(r)\in\mathbb{C}^{8},\label{47.2}\\
F_{1}(r)  &  :=\sum\limits_{k=0,1,2}\frac{\exp\left\{  -i\sqrt{2}\left\Vert
r\right\Vert _{\mathbb{R}^{8}}\sin(\phi(r)+2\pi k/3)\right\}  }{1-2\cos
(2(\phi(r)+2\pi k/3))},\nonumber\\
F_{2}\left(  r\right)   &  :=\frac{2}{\sqrt{3}}\sum\limits_{k=0,1,2}%
\sin\left(  \phi(r)+2\pi k/3\right) \nonumber\\
&  \times\frac{\exp\left\{  -i\sqrt{2}\left\Vert r\right\Vert _{\mathbb{R}%
^{8}}\sin(\phi(r)+2\pi k/3)\right\}  }{1-2\cos(\left(  2(\phi(r)+2\pi
k/3)\right)  }.\nonumber
\end{align}
Taking into account (\ref{36}), (\ref{47}), (\ref{47.1}) we derive that, for
any vector $r\in\mathbb{R}^{8},$%
\begin{equation}
\exp\left\{  -i\sqrt{\frac{3}{2}}\text{ }(r\cdot\Lambda)\right\}
=\frac{\mathbb{I}}{3}K_{3}(r)+\left\{  \text{ }F_{1}\left(  \phi(r)\right)
\text{ }\left(  p(r)\cdot\Lambda\right)  +F_{2}\left(  \phi(r)\right)  \text{
}(r\cdot\Lambda)\text{ }\right\}  \label{47.3}%
\end{equation}
In view of relations (\ref{9}) and (\ref{47.2}), this expression can be
represented otherwise in the form%
\begin{align}
\exp\left\{  -i\sqrt{\frac{3}{2}}\text{ }(r\cdot\Lambda)\right\}   &
=\sum_{k=0,1,2}{\LARGE \{[}\frac{1}{\left\Vert r\right\Vert _{\mathbb{R}^{8}%
}^{2}}(r\cdot\Lambda)^{2}+\frac{2}{\sqrt{3}\left\Vert r\right\Vert
_{\mathbb{R}^{8}}}(r\cdot\Lambda)\sin\left(  \phi(r)+2\pi k/3\right)
\label{47.4}\\
&  -\frac{\mathbb{I}}{3}[1+2\cos\left(  2(\phi(r)+2\pi k/3)\right)
]{\LARGE \}}\nonumber\\
&  \times\frac{\exp\left\{  -i\sqrt{2}\left\Vert r\right\Vert _{\mathbb{R}%
^{8}}\sin\left(  \phi(r)+2\pi k/3\right)  \right\}  }{1-2\cos\left(
2(\phi(r)+2\pi k/3)\right)  },\nonumber
\end{align}
which agrees with formula (\ref{4}) in \cite{Curt}.
\end{itemize}

\section{General nonstationary qubit case}

In this section, based on the new general results derived in Sections 2 and 3,
we specify the unitary evolution operator (\ref{39}) for $d=2.$

In the qubit case, $\Lambda\equiv\sigma=(\sigma_{1},\sigma_{2},\sigma_{3})$
and a general Hamiltonian on $\mathbb{C}^{2}$ has the form
\begin{equation}
H(t)=b_{0}(t)\mathbb{I}+b(t)\cdot\sigma,\text{ \ \ \ }b(t)\in\mathbb{R}^{3}.
\label{49}%
\end{equation}
Here and in what follows, for short, we suppress the lower index $H$ in
notations $n_{H}(t)$, $b_{H}(t)\in\mathbb{R}^{3}$ and the lower index
$\mathbb{R}^{3}$ in notation $\left\Vert \cdot\right\Vert _{\mathbb{R}^{3}}.$

Let us specify the main issues of Theorem 1 if $d=2.$ In this case:

\begin{itemize}
\item The structure coefficients $d_{kmj}=0$, for all $k,m,j=1,2,3$, and
coefficients $f_{kmj}=\epsilon_{kmj}$ constitute the Levi-Civita symbol.
Therefore, the system of linear ODEs (\ref{25}) reduces to
\begin{align}
\overset{\cdot}{u}_{0}(t,t_{0})  &  =b(t)\cdot\widetilde{u}(t,t_{0}),\text{
\ \ \ }\widetilde{u}_{0}(t_{0},t_{0})=1,\label{50}\\
\overset{\cdot}{\widetilde{u}}(t,t_{0})  &  =-u_{0}(t,t_{0})b_{j}%
+b(t)\times\widetilde{u}(t,t_{0}),\text{ \ \ \ }\widetilde{u}(t_{0}%
,t_{0})=0,\nonumber\\
(u_{0}(t,t_{0}))^{2}+\left\Vert \widetilde{u}(t,t_{0})\right\Vert
_{\mathbb{R}^{3}}^{2}  &  =1,\nonumber
\end{align}
with $u_{0}(t,t_{0})\in\mathbb{R}$, $\widetilde{u}(t,t_{0})\in\mathbb{R}^{3}$
and notation $b\times\widetilde{u}\ $for a vector product on $\mathbb{R}^{3}$.

By introducing a 4-dimensional real-valued unit vector $q(t,t_{0}%
)=(u_{0}(t,t_{0}),\widetilde{u}(t,t_{0}))\in\mathbb{R}^{4}$ and denoting by
$q^{\prime}(t,t_{0})$ the column-vector with elements comprised of components
of vector $q(t,t_{0}),$ we rewrite the system of linear ODEs (\ref{50}) in the
normal form%
\begin{equation}
\frac{d}{dt}q^{\prime}(t,t_{0})=\mathrm{A}(t)q^{\prime}(t,t_{0}),\text{
\ \ }q(t,t_{0})=(1,0,0,0), \label{51}%
\end{equation}
with the skew-symmetric matrix
\begin{equation}
\mathrm{A}(t)=%
\begin{pmatrix}
0 & b_{1}(t) & b_{2}(t) & b_{3}(t)\\
-b_{1}(t) & 0 & -b_{3}(t) & b_{2}(t)\\
-b_{2}(t) & b_{3}(t) & 0 & -b_{1}(t)\\
-b_{3}(t) & -b_{2}(t) & b_{1}(t) & 0
\end{pmatrix}
. \label{52}%
\end{equation}

\item For $d=2$, function (\ref{34}) and its gradient are given due to
(\ref{44}) by $K_{2}(n)=\cos\left\Vert n(t)\right\Vert ,$ $\triangledown
_{n}K_{2}(n)=-2\sin\left(  \left\Vert n\right\Vert \right)  \frac
{n}{\left\Vert n\right\Vert },$ so that by (\ref{40})
\begin{equation}
u_{0}(t,t_{0})=\cos\left\Vert n(t,t_{0})\right\Vert \in\mathbb{R}%
,\ \ \ \widetilde{u}(t,t_{0})=-\sin\left(  \left\Vert n(t,t_{0})\right\Vert
\right)  \frac{n(t,t_{0})}{\left\Vert n(t,t_{0})\right\Vert }\in\mathbb{R}%
^{3}, \label{52.2}%
\end{equation}

and the first and the second equations in (\ref{50}) take the forms%
\begin{equation}
\frac{d\left\Vert n(t,t_{0})\right\Vert }{dt}=\frac{b(t)\cdot n(t,t_{0}%
)}{\left\Vert n(t)\right\Vert } \label{52.3}%
\end{equation}
and%
\begin{align}
&  \left(  \frac{\sin\left\Vert n(t,t_{0})\right\Vert }{\left\Vert
n(t,t_{0})\right\Vert }-\cos\left\Vert n(t)\right\Vert \right)  \frac
{d\left\Vert n(t,t_{0})\right\Vert }{dt}\frac{n(t,t_{0})}{\left\Vert
n(t,t_{0})\right\Vert }-\frac{\sin\left\Vert n(t,t_{0})\right\Vert
}{\left\Vert n(t,t_{0})\right\Vert }\frac{dn(t,t_{0})}{dt}\label{52.4}\\
&  =-b\cos\left\Vert n(t,t_{0})\right\Vert +\frac{b(t)\times n(t,t_{0}%
)}{\left\Vert n(t,t_{0})\right\Vert }\sin\left\Vert n(t,t_{0})\right\Vert
,\nonumber
\end{align}
respectively.

\item The Cauchy problem (\ref{41}) in Theorem 1 reduces to%
\begin{align}
\frac{dn(t,t_{0})}{dt}  &  =b(t)+n_{\bot}(t,t_{0})\ \text{\ \ }t>t_{0}%
,\label{53}\\
n(t_{0},t_{0})  &  =0,\nonumber
\end{align}
where vector $n_{\bot}(t)\in\mathbb{R}^{3}$ is orthogonal for all $t>t_{0}$ to
vector $n(t)\in\mathbb{R}^{3}$ and is determined via Eq. (\ref{42}). For
$d=2$, the latter equation reduces to
\begin{align}
&  \left(  \frac{\sin\left\Vert n(t,t_{0})\right\Vert }{\left\Vert
n(t,t_{0})\right\Vert }-\cos\left\Vert n(t,t_{0})\right\Vert \right)  \left(
\frac{b(t)\cdot n(t,t_{0})}{\left\Vert n(t,t_{0})\right\Vert ^{2}}%
n(t,t_{0})\text{ }-\text{ }b(t)\right) \label{54}\\
&  =\frac{\sin\left\Vert n(t,t_{0})\right\Vert }{\left\Vert n(t,t_{0}%
)\right\Vert }n_{\perp}(t,t_{0})\text{ }+\text{ }\left(  b(t)\times
n(t,t_{0})\right)  \frac{\sin\left\Vert n(t,t_{0})\right\Vert }{\left\Vert
n(t,t_{0})\right\Vert }.\nonumber
\end{align}
Noting that, on the left-hand side of (\ref{54}),
\begin{equation}
\frac{b\cdot n}{\left\Vert n\right\Vert ^{2}}n\text{ \ }-\text{ }b=-\frac
{1}{\left\Vert n\right\Vert ^{2}}\left(  n\times b\times n\right)  ,
\label{55}%
\end{equation}
and vectors
\begin{equation}
n\times b\times n,\text{ \ \ \ }b\times n \label{56}%
\end{equation}
are mutually orthogonal and are both in the plane orthogonal to vector
$n(t,t_{0})\in\mathbb{R}^{3},$ we represent vector $n_{\perp}(t,t_{0})$ in
(\ref{53}) and (\ref{54}) as%
\begin{equation}
n_{\perp}(t,t_{0})=\alpha(t)\left(  b(t)\times n(t,t_{0})\right)  \text{
}+\text{ }\beta(t,t_{0})\left(  \text{ }n(t,t_{0})\times b(t)\times
n(t,t_{0})\right)  \label{57}%
\end{equation}
and find via (\ref{54}) that%
\begin{equation}
\alpha(t,t_{0})=-1,\text{ \ \ \ \ \ }\beta(t,t_{0})=-\frac{1-\left\Vert
n(t,t_{0})\right\Vert \mathrm{ctg}\left\Vert n(t,t_{0})\right\Vert
}{\left\Vert n(t,t_{0})\right\Vert ^{2}}. \label{58}%
\end{equation}
Therefore, Eqs. (\ref{53})--(\ref{58}) imply
\begin{align}
\frac{dn(t,t_{0})}{dt}  &  =b(t)-\left(  b(t)\times n(t,t_{0})\right)
\label{59}\\
&  -\text{ }\frac{1-\left\Vert n(t,t_{0})\right\Vert \mathrm{ctg}\left\Vert
n(t,t_{0})\right\Vert }{\left\Vert n(t,t_{0})\right\Vert ^{2}}\left(
n(t,t_{0})\times b(t)\times n(t,t_{0})\right)  ,\nonumber\\
n(t_{0},t_{0})  &  =0.\nonumber
\end{align}
\bigskip
\end{itemize}

Theorem 1 and relations (\ref{50})--(\ref{59}) prove the following statement
on the unitary evolution of a qubit in a general nonstationary case.

\begin{theorem}
Let $H(t)=b_{0}(t)\mathbb{I}+b(t)\cdot\sigma,$ $b(t)\in\mathbb{R}^{3},$ be a
qubit Hamiltonian on $\mathbb{C}^{2}$. The unitary operator $U_{H}(t,t_{0})$
on $\mathbb{C}^{2}$describing the evolution of a qubit under Hamiltonian
$H(t)$ takes the form
\begin{align}
U_{H}(t,t_{0})  &  =\exp\left\{  -i\int_{t_{0}}^{t}b_{0}(\tau)d\tau\right\}
\exp\left\{  -i\text{ }\left(  n(t,t_{0})\cdot\sigma\right)  \right\}
\label{60}\\
&  =\exp\left\{  -i\int_{t_{0}}^{t}b_{0}(\tau)d\tau\right\}  \left(  \text{
}u_{0}(t,t_{0})\text{ }\mathbb{I}\text{ }+\text{ }i\text{ }\widetilde{u}%
(t,t_{0})\cdot\sigma\right)  ,\nonumber
\end{align}
where the unit vector $\left(  u_{0}(t,t_{0}),\widetilde{u}(t,t_{0})\right)
\in\mathbb{R}^{4}$ is the solution of the Cauchy problem (\ref{50})
(equivalently, (\ref{51})), vector $n(t,t_{0})\in\mathbb{R}^{3}$ is the
solution of the Cauchy problem (\ref{59}) and the following relations hold%
\begin{align}
u_{0}(t,t_{0})  &  =\cos\left\Vert n(t,t_{0})\right\Vert \in\mathbb{R}%
,\ \ \ \widetilde{u}(t,t_{0})=-\sin\left(  \left\Vert n(t,t_{0})\right\Vert
\right)  \frac{n(t,t_{0})}{\left\Vert n(t,t_{0})\right\Vert },\label{61}\\
\frac{n(t,t_{0})}{\left\Vert n(t,t_{0})\right\Vert }  &  =-\frac
{\text{\ }\widetilde{u}(t,t_{0})}{\text{\ }\left\Vert \widetilde{u}%
(t,t_{0})\right\Vert },\text{ \ \ }\left\Vert n(t,t_{0})\right\Vert
=\arccos\left(  u_{0}(t,t_{0})\right)  .\nonumber
\end{align}

\end{theorem}

The cocycle property (\ref{2.1}) implies that, in the qubit case, the unit
vector $\left(  u_{0}(t,t_{0}),\widetilde{u}(t,t_{0})\right)  \in
\mathbb{R}^{4}$ in (\ref{60}) -- \ which is solution of the Cauchy problem
(\ref{51}) -- must satisfy the relations
\begin{align}
u_{0}(t,s)u_{0}(s,t_{0})\text{ }-\text{ }\widetilde{u}(t,s)\cdot
\widetilde{u}(s,t_{0})  &  =u_{0}(t,t_{0}),\label{61.1}\\
u_{0}(t,s)\widetilde{u}(s,t_{0})+u_{0}(s,t_{0})\widetilde{u}(t,s)\text{
}-\text{ }\widetilde{u}(t,s)\times\widetilde{u}(s,t_{0})  &  =\widetilde{u}%
(t,t_{0}).\nonumber
\end{align}

For $d=2$, relation (\ref{13}) reduces to the condition%
\begin{equation}
b(t)\times\left(  \int_{t_{0}}^{t}b(\tau)d\tau\right)  =0, \label{62}%
\end{equation}
which is necessary and sufficient for the Cauchy problem (\ref{50})
(equivalently, (\ref{51})) and the Cauchy problem (\ref{59}) to have the
solutions
\begin{align}
n(t,t_{0})  &  =\int_{t_{0}}^{t}b(\tau)d\tau,\text{ \ \ \ \ }u_{0}%
(t,t_{0})=\cos\left(  \left\Vert \int_{t_{0}}^{t}b(\tau)d\tau\right\Vert
\right)  ,\label{63}\\
\widetilde{u}(t,t_{0})  &  =-\frac{\sin\left(  \left\Vert \int_{t_{0}}%
^{t}b(\tau)d\tau\right\Vert \right)  }{\left\Vert \int_{t_{0}}^{t}b(\tau
)d\tau\right\Vert }\left(  \int_{t_{0}}^{t}b(\tau)d\tau\right)  ,\nonumber
\end{align}
and the unitary evolution operator $U_{H}(t,t_{0})$ to be given by
\begin{align}
U_{H}(t,t_{0})  &  =\exp\left\{  -i\int_{t_{0}}^{t}b_{0}(\tau)d\tau\right\}
\exp\left\{  -i\text{ }\left(  \int_{t_{0}}^{t}\left(  b(\tau)\cdot
\sigma\right)  d\tau\right)  \right\} \label{64}\\
&  =\exp\left\{  -i\int_{t_{0}}^{t}b_{0}(\tau)d\tau\right\}  \left[
\mathbb{I}\cos\left(  \left\Vert \int_{t_{0}}^{t}b(\tau)d\tau\right\Vert
\right)  \mathbb{-}\text{ }i\frac{\sin\left(  \left\Vert \int_{t_{0}}%
^{t}b(\tau)d\tau\right\Vert \right)  }{\left\Vert \int_{t_{0}}^{t}b(\tau
)d\tau\right\Vert }\left(  \int_{t_{0}}^{t}b(\tau)d\tau\cdot\sigma\right)
\right]  .\nonumber
\end{align}
The expression standing in the first line of (\ref{64}) is consistent with
expression (\ref{14}) valid under the general qudit condition (\ref{5}) and
specified for $d=2$.

Condition (\ref{62}) is, in particular, true if $b(t)=e_{b}\left\Vert
b(t)\right\Vert $ where a unit vector $e_{b}$ does not vary in time.
Substituting this $b(t)$ into (\ref{64}) we have%
\begin{align}
U_{H}(t,t_{0})  &  =\exp\left\{  -i\int_{t_{0}}^{t}b_{0}(\tau)d\tau\right\}
\exp\left\{  -i\text{ }\left(  e_{b}\cdot\sigma\right)  \int_{t_{0}}%
^{t}\left\Vert b(\tau)\right\Vert \mathrm{d}\tau\right\} \label{65}\\
&  =\exp\left\{  -i\int_{t_{0}}^{t}b_{0}(\tau)d\tau\right\}  \left[
\mathbb{I}\cos\left(  \int_{t_{0}}^{t}\left\Vert b(\tau)\right\Vert
\mathrm{d}\tau\right)  \text{ }-\text{ }i\sin\left(  \int_{t_{0}}%
^{t}\left\Vert b(\tau)\right\Vert \mathrm{d}\tau\right)  \left(  e_{b}%
\cdot\sigma\right)  \right]  {\LARGE .}\nonumber
\end{align}

In the following section, based on the general result formulated in Theorem 2,
we specify classes of nonstationary Hamiltonians $H(t),$ for which we can find
the precise solutions of the Cauchy problem (\ref{50}) (equivalently,
(\ref{51})) and, hence, explicitly specify the unitary operator (\ref{60}) via
coefficients $b_{0}(t),$ $b(t)$ of a Hamiltonian $H(t).$

\section{Special classes of qubit Hamiltonians}

Let, for a qubit Hamiltonian (\ref{49}), components $(\left\Vert
b(t)\right\Vert ,\theta_{b}(t),\varphi_{b}(t))$ of a vector $b(t)\in
\mathbb{R}^{3}$ in the spherical coordinate system be such that\footnote{Here,
we suppose that $b(t)$ is twice differentiable.}%
\begin{align}
\frac{d}{dt}J_{1}  &  =0,\text{ \ \ where }J_{1}:=\frac{1}{\Omega_{b}%
(t)}\left(  \cos\left(  \theta_{b}(t)\right)  -\frac{\overset{\cdot}{\varphi
}_{b}(t)}{2\left\Vert b(t)\right\Vert }\right)  ,\label{66}\\
\frac{d}{dt}J_{2}  &  =0,\text{ \ \ where }J_{2}:=\frac{\sin\left(  \theta
_{b}(t)\right)  }{\Omega_{b}(t)},\nonumber
\end{align}
where
\begin{equation}
\Omega_{b}(t):=\sqrt{\left(  \cos\left(  \theta_{b}(t)\right)  -\frac
{\overset{\cdot}{\varphi_{b}}(t)}{2\left\Vert b(t)\right\Vert }\right)
^{2}+\sin^{2}\left(  \theta_{b}(t)\right)  }, \label{67}%
\end{equation}
so that $J_{1}^{2}+J_{2}^{2}=1.$

The class of Hamiltonians specified by conditions (\ref{66}) is rather broad
and includes, in particular, all cases studied in the literature for which:%
\begin{equation}
\overset{\cdot}{\theta}_{b}(t)=0,\text{ \ \ }\overset{\cdot\cdot}{\varphi_{b}%
}(t)=0. \label{68}%
\end{equation}
Represented otherwise, constant $J_{1}$ takes the form
\begin{align}
J_{1}  &  =\frac{1}{\left\Vert b(t)\right\Vert \Omega_{b}(t)}\left(
b_{3}(t)-\frac{1}{2}\frac{d}{dt}\varphi_{b}(t)\right)  ,\label{69}\\
\left\Vert b(t)\right\Vert \Omega_{b}(t)  &  =\sqrt{\left(  b_{3}(t)-\text{
}\frac{1}{2}\frac{d}{dt}\varphi_{b}(t)\right)  ^{2}+b_{1}^{2}(t)+b_{2}^{2}%
(t)},\nonumber\\
\mathrm{tg}(\varphi_{b}(t))  &  =b_{2}(t)/b_{1}(t),\nonumber
\end{align}
from which it is immediately clear that the class of Hamiltonians specified by
conditions (\ref{66}) is defined via the special time behavior of a vector
$b(t)$ with respect to the $\mathrm{x}_{3}$-axis.

Quite similarly we can introduce the class of Hamiltonians specified via the
property of $b(t)\in\mathbb{R}^{3}$ which is similar by its form to (\ref{69})
but with respect to the \textrm{x}$_{1}$-axis or the \textrm{x}$_{2}$-axis.

Though, in the following statement, we explicitly specify only the unitary
qubit evolution (\ref{60}) under a Hamiltonian satisfying conditions
(\ref{66}), the new result of this statement can be easily reformulated for
the classes of nonstationary Hamiltonians specified by conditions on
$b(t)\in\mathbb{R}^{3}$ with respect to the \textrm{x}$_{1}$-axis and the
\textrm{x}$_{2}$-axis.

\begin{theorem}
Let, for a qubit Hamiltonian $H(t)=b_{0}(t)\mathbb{I}+b(t)\cdot\sigma$ on
$\mathbb{C}^{2}$ the conditions (\ref{66}) be fulfilled. Then, for the unitary
operator $U_{H}(t,t_{0})$ given by relations (\ref{60}), (\ref{61}) and
describing the evolution of a qubit state under a Hamiltonian $H(t)$, the unit
vector $(u_{0}(t,t_{0}),\widetilde{u}(t,t_{0}))\in\mathbb{R}^{4}$ -- the
solution of the Cauchy problem (\ref{50}), equivalently, (\ref{51}), takes the
form%
\begin{align}
u_{0}(t,t_{0})  &  =\cos\left(  \frac{\varphi_{b}(t)-\varphi_{b}(t_{0})}%
{2}\right)  \cos\left(  \gamma_{b}(t,t_{0})\right)  -J_{1}\sin\left(
\frac{\varphi_{b}(t)-\varphi_{b}(t_{0})}{2}\right)  \sin\left(  \gamma
_{b}(t,t_{0})\right)  ,\label{70}\\
\widetilde{u}_{1}(t,t_{0})  &  =-J_{2}\cos\left(  \frac{\varphi_{b}%
(t)+\varphi_{b}(t_{0})}{2}\right)  \sin\left(  \gamma_{b}(t,t_{0})\right)
,\nonumber\\
\widetilde{u}_{2}(t,t_{0})  &  =-J_{2}\sin\left(  \frac{\varphi_{b}%
(t)+\varphi_{b}(t_{0})}{2}\right)  \sin\left(  \gamma_{b}(t,t_{0})\right)
,\nonumber\\
\widetilde{u}_{3}(t,t_{0})  &  =-J_{1}\cos\left(  \frac{\varphi_{b}%
(t)-\varphi_{b}(t_{0})}{2}\right)  \sin\left(  \gamma_{b}(t,t_{0})\right)
-\sin\left(  \frac{\varphi_{b}(t)-\varphi_{b}(t_{0})}{2}\right)  \cos\left(
\gamma_{b}(t,t_{0})\right)  ,\nonumber
\end{align}
satisfying the cocycle property (\ref{61.1}). In (\ref{70}),
\begin{equation}
\gamma_{b}(t,t_{0}):=\int_{t_{0}}^{t}\left\Vert b(\tau)\right\Vert \Omega
_{b}(\tau)\text{ }\mathrm{d}\tau\label{71}%
\end{equation}
and $\theta_{b}(t),$ $\varphi_{b}(t)$ are angles specifying at time $t$ vector
$b(t)\in\mathbb{R}^{3}$ in the spherical coordinate system.
\end{theorem}

The proof of this statement is given in Appendix A. Note that the Cauchy
problem with a skew-symmetric matrix -- like the one in (\ref{51})-- arises in
many fields of mathematical physics, for example, in the solid body theory, in
the quaternions models \cite{Cheln}, etc. If we reformulate conditions
(\ref{69}) (equivalently, (\ref{66})) with respect to the \textrm{x}$_{1}%
$-axis, then the corresponding solution $(u_{0}(t,t_{0}),\widetilde{u}%
(t,t_{0}))\in\mathbb{R}^{4}$ of the Cauchy problem for the ODEs (\ref{51})
would agree with the treatment in Section 5.10 of Ref. \cite{Cheln}.

Let, for example, $b(t)=e_{b}\left\Vert b(t)\right\Vert $ where a unit vector
$e_{b}$ does not vary in time -- the case we have analyzed above in (\ref{65})
and where the general condition (\ref{62}) is true. In this case,
\begin{align}
\varphi_{b}(t)  &  =\varphi_{b}(t_{0})=\varphi_{b},\text{ \ \ \ }\theta
_{b}(t)=\theta_{b}(t_{0})=\theta_{b},\label{72}\\
J_{1}  &  =\cos\theta_{b},\text{ \ \ }J_{2}=\sin\theta_{b},\nonumber
\end{align}
conditions (\ref{66}) are also fulfilled and the substitution of (\ref{72})
into expression (\ref{69}) leads exactly to relation (\ref{64}).

However, in general, conditions (\ref{62}) and (\ref{66}) do not need to be
fulfilled simultaneously.

As an application of the result of Theorem 3, consider some examples important
for applications where conditions (\ref{66}) are fulfilled while condition
(\ref{62}) is violated.

\begin{enumerate}
\item Let, for a qubit Hamiltonian (\ref{49}), the spherical coordinates of a
vector $b(t)\in\mathbb{R}^{3}$ satisfy the relations%
\begin{equation}
\theta_{b}(t)=\theta_{b},\text{ \ \ }\varphi_{b}(t)=\omega t+\eta,\text{
\ \ }\left\Vert b(t)\right\Vert =b,\text{\ \ \ }\eta\in\mathbb{R}, \label{73}%
\end{equation}
-- the case where a vector $b(t)$ rotates around the $\mathrm{x}_{3}$-axis
with an angular velocity $\omega$ and has a norm constant in time. Based on
approaches different to ours, this case was considered in many papers in
connection with the evolution of a pure qubit state, see for example, in
\cite{Flu}. For case (\ref{73}), conditions (\ref{66}) and parameters in
(\ref{70}) take the forms:%
\begin{align}
J_{1}  &  =\frac{\cos\theta_{b}-\omega/2b}{\Omega_{b}},\text{ \ \ \ }%
J_{2}=\frac{\sin\theta_{b}}{\Omega_{b}},\label{74}\\
\Omega_{b}  &  =\sqrt{\left(  \cos\theta_{b}-\omega/2b\right)  ^{2}+\sin
^{2}\theta_{b}}=Const,\nonumber\\
\left\Vert b(t)\right\Vert \Omega_{b}  &  =\sqrt{\left(  2b\cos\theta
_{b}-\omega\right)  ^{2}+4b^{2}\sin^{2}\theta_{b}}:=\widetilde{\Omega}%
_{b}.\nonumber
\end{align}
Therefore, for case (\ref{73}) we have by Theorem 3:%
\begin{align}
u_{0}(t,t_{0})  &  =\cos\left(  \frac{\omega(t-t_{0})}{2}\right)
\cos(\widetilde{\Omega}_{b}(t-t_{0}))-J_{1}\sin\left(  \frac{\omega(t-t_{0}%
)}{2}\right)  \sin(\widetilde{\Omega}_{b}(t-t_{0})),\nonumber\\
\widetilde{u}_{1}(t,t_{0})  &  =-J_{2}\cos\left(  \frac{\omega(t+t_{0})}%
{2}+\eta\right)  \sin(\widetilde{\Omega}_{b}(t-t_{0})),\label{75}\\
\widetilde{u}_{2}(t,t_{0})  &  =-J_{2}\sin\left(  \frac{\omega(t+t_{0})}%
{2}+\eta\right)  \sin(\widetilde{\Omega}_{b}(t-t_{0})),\nonumber\\
\widetilde{u}_{3}(t,t_{0})  &  =-J_{1}\cos\left(  \frac{\omega(t-t_{0})}%
{2}\right)  \sin(\widetilde{\Omega}_{b}(t-t_{0}))-\sin\left(  \frac
{\omega(t-t_{0})}{2}\right)  \cos(\widetilde{\Omega}_{b}(t-t_{0})),\nonumber
\end{align}
so that the unitary evolution operator (\ref{60}) with the unit vector
$(u_{0}(t),\widetilde{u}(t))$ given by (\ref{75}) completely defines the
evolution of every qubit state under a nonstanionary Hamiltonian specified by
relations (\ref{73}).

Taking, for example, $t_{0}=0$ and an initial pure state $|\Psi(0)\rangle
=|0\rangle\in\mathbb{C}^{2}$ we find that at any moment $t>0$ the pure state
is
\begin{align}
|\Psi(t)\rangle &  =U_{H}(t)|0\rangle=u_{0}(t,0)|0\rangle+i\widetilde{u}%
_{1}(t,0)|1\rangle-\widetilde{u}_{2}(t,0)|1\rangle+i\widetilde{u}%
_{3}(t,0)|0\rangle\label{76}\\
&  =\left(  u_{0}(t,0)+i\widetilde{u}_{3}(t,0)\right)  |0\rangle+i\left(
\widetilde{u}_{1}(t,0)+i\widetilde{u}_{2}(t,0)\right)  |1\rangle,\nonumber
\end{align}
where $|0\rangle,|1\rangle$ are elements of the computational basis of
$\mathbb{C}^{2}.$ Substituting (\ref{75}) into (\ref{76}), we have
\begin{align}
u_{0}(t,0)+i\widetilde{u}_{3}(t,0)  &  =\left(  \cos\left(  \widetilde{\Omega
}_{b}t\right)  -iJ_{1}\sin\left(  \widetilde{\Omega}_{b}t\right)  \right)
\exp\left\{  -\frac{i\omega t}{2}\right\}  ,\label{77}\\
\widetilde{u}_{1}(t,0)+i\widetilde{u}_{2}(t,0)  &  =-J_{2}\sin
(\widetilde{\Omega}_{b}t)\exp\left\{  \frac{i\omega t}{2}+\eta\right\}
,\nonumber
\end{align}
so that%
\begin{align}
|\Psi(t)\rangle &  =\left\{  \cos\left(  \widetilde{\Omega}_{b}t\right)
-iJ_{1}\sin\left(  \widetilde{\Omega}_{b}t\right)  \right\}  \exp\left\{
-\frac{i\omega t}{2}\right\}  |0\rangle\label{77.1}\\
&  -iJ_{2}\sin(\widetilde{\Omega}_{b}t)\exp\left\{  \frac{i\omega t}{2}%
+\eta\right\}  |1\rangle.\nonumber
\end{align}
where constants $J_{1}$ and $J_{2}$ are given by (\ref{74}). For $\eta=0$, the
pure state (\ref{77}) coincides with the pure state given by Eq. (138.11) in
Ref. \cite{Flu} and found by another approach.

\item Consider further a more general case, where, for a vector $b(t)\in
\mathbb{R}^{3}$ in (\ref{49}):
\begin{align}
b_{1}(t)  &  =q\frac{\overset{\cdot}{\varphi}_{b}(t)}{\lambda}\cos\left(
\varphi_{b}(t)\right)  ,\text{ \ \ }b_{2}(t)=q\frac{\overset{\cdot}{\varphi
}_{b}(t)}{\lambda}\sin\left(  \varphi_{b}(t)\right)  ,\label{78}\\
b_{3}(t)  &  =p\frac{\overset{\cdot}{\varphi}_{b}(t)}{\lambda},\nonumber
\end{align}
with function$\frac{\overset{\cdot}{\varphi}_{b}(t)}{\lambda}>0$ for all
$t>t_{0}$ and some constants $\lambda,$ $q,$ $p$. In this case,
\begin{align}
\left\Vert b(t)\right\Vert  &  =\frac{\overset{\cdot}{\varphi}_{b}(t)}%
{\lambda}\sqrt{q^{2}+p^{2}},\text{ \ \ }\cos(\theta_{b}(t))=\frac{p}%
{\sqrt{q^{2}+p^{2}}}=Const,\label{79}\\
\Omega_{b}(t)  &  =\frac{1}{\sqrt{q^{2}+p^{2}}}\sqrt{\left(  p-\frac{\lambda
}{2}\right)  ^{2}+q^{2}}=Const,\nonumber\\
\left\Vert b(t)\right\Vert \Omega_{b}  &  =\frac{\overset{\cdot}{\varphi}%
_{b}(t)}{\lambda}\sqrt{\left(  p-\frac{\lambda}{2}\right)  ^{2}+q^{2}}%
=\frac{\zeta}{\lambda}\overset{\cdot}{\varphi}_{b}(t),\nonumber\\
\zeta &  :=\sqrt{\left(  p-\frac{\lambda}{2}\right)  ^{2}+q^{2}}%
=Const\nonumber
\end{align}
Hence, by (\ref{66}) the constants
\begin{equation}
J_{1}=\frac{p-\lambda/2}{\zeta},\text{ \ \ \ }J_{2}=\frac{q}{\zeta},
\label{79.1}%
\end{equation}
and, in Theorem 3, the vector $(u_{0}(t),\widetilde{u}(t))\in\mathbb{R}^{4},$
which specifies by (\ref{60}) the unitary evolution of a qubit, is given by%
\begin{align}
u_{0}(t,t_{0})  &  =\cos\left(  \frac{\varphi_{b}(t)-\varphi_{b}(t_{0})}%
{2}\right)  \cos\left\{  \frac{\zeta}{\lambda}(\varphi_{b}(t)-\varphi
_{b}(t_{0}))\right\} \label{81}\\
&  -\frac{p-\lambda/2}{\zeta}\sin\left(  \frac{\varphi_{b}(t)-\varphi
_{b}(t_{0})}{2}\right)  \sin\left\{  \frac{\zeta}{\lambda}(\varphi
_{b}(t)-\varphi_{b}(t_{0}))\right\}  ,\nonumber\\
\widetilde{u}_{1}(t,t_{0})  &  =-\frac{q}{\zeta}\cos\left(  \frac{\varphi
_{b}(t)+\varphi_{b}(t_{0})}{2}\right)  \sin\left\{  \frac{\zeta}{\lambda
}(\varphi_{b}(t)-\varphi_{b}(t_{0})\right\}  ,\nonumber\\
\widetilde{u}_{2}(t,t_{0})  &  =-\frac{q}{\zeta}\sin\left(  \frac{\varphi
_{b}(t)+\varphi_{b}(t_{0})}{2}\right)  \sin\left\{  \frac{\zeta}{\lambda
}(\varphi_{b}(t)-\varphi_{b}(t_{0}))\right\}  ,\nonumber\\
\widetilde{u}_{3}(t,t_{0})  &  =-\frac{p-\lambda/2}{\zeta}\cos\left(
\frac{\varphi_{b}(t)-\varphi_{b}(t_{0})}{2}\right)  \sin\left\{  \frac{\zeta
}{\lambda}(\varphi_{b}(t)-\varphi_{b}(t_{0}))\right\} \nonumber\\
&  +\sin\left(  \frac{\varphi_{b}(t)-\varphi_{b}(t_{0})}{2}\right)
\cos\left\{  \frac{\zeta}{\lambda}(\varphi_{b}(t)-\varphi_{b}(t_{0}))\right\}
,\nonumber
\end{align}
where $\lambda,$ $q,$ $p$ are some constants and angle $\varphi_{b}(t)$ is an
arbitrary function of $t$, such that $\frac{\overset{\cdot}{\varphi}_{b}%
(t)}{\lambda}>0$. \ If, in particular, $\overset{\cdot}{\varphi}_{b}%
(t)=\omega$ and $\lambda=\omega,$ then relations (\ref{81}) reduce to
relations (\ref{75}).
\end{enumerate}

\section{Conclusions}

In the present article, we introduced a new general formalism that allows for
the analysis of the unitary evolution of a qudit ($d\geq2$) under an arbitrary
time-dependent Hamiltonian $H(t)$ in terms of the Bloch-like vector space. Via
this formalism we derived (Theorem 1, Section 3) the new general equations
specifying the evolution of the Bloch-like vector in the generalized Gell-Mann
representation of $U_{H}(t,t_{0})$ and the vector $n(t,t_{0})\in
\mathbb{R}^{d^{2}-1}$ in the exponential representation of $U_{H}(t,t_{0})$.

Applying the general equations (\ref{25}), (\ref{41}), (\ref{42}) to a qubit
case ($d=2$), we then derived (Theorem 2, Section 4) a new general result on
the qubit evolution under a nonstationary Hamiltonian. This general result
allowed us to find (Theorem 3, Section 5) the new precise analytical solutions
for a wide class of nonstationary Hamiltonians which comprise the qubit cases
already known in the literature only as particular ones.

The general formalism presented in this article is valid for a qudit of an
arbitrary dimension $d>2,$ in particular, for a qutrit and the analysis of the
evolution of a qutrit under a time-dependent Hamiltonian within this new
formalism is a subject of our future research.

\section*{Acknowledgments}

E.R.L. is grateful to Professor A. Khrennikov and Professor A. Borisov for
useful discussions. The study by E.R.L. in Sections 2 and 3 is supported by
the Russian Science Foundation under the grant No 19-11-00086 and performed at
the Steklov Mathematical Institute of Russian Academy of Sciences. The study
by E.R.L. and C.K. in Sections 4 and 5 is performed at the National Research
University Higher School of Economics.

\section{Appendix A}

In this section, we present the proof of Theorem 3, namely, we show that
functions $u_{0}(t)\in\mathbb{R},$ $\widetilde{u}(t)\in\mathbb{R}^{3},$ given
by (\ref{70}), constitute solutions of the Cauchy problem (\ref{50}),
equivalently (\ref{51}), under conditions (\ref{66}) and satisfy also the
cocycle property (\ref{61.1}).

Under conditions (\ref{66}), the derivative of function $u_{0}(t)\in
\mathbb{R}$ in (\ref{70}) has the form%
\begin{align}
\frac{d}{dt}u_{0}(t)  &  =-\frac{1}{2}\frac{d\varphi_{b}(t)}{dt}\sin\left(
\frac{\varphi_{b}(t)-\varphi_{b}(t_{0})}{2}\right)  \cos\left(  \gamma
_{b}(t,t_{0})\right) \tag{A1}\\
&  -\Vert b(t)\Vert\Omega_{b}(t)\cos\left(  \frac{\varphi_{b}(t)-\varphi
_{b}(t_{0})}{2}\right)  \sin\left(  \gamma_{b}(t,t_{0})\right) \nonumber\\
&  -\frac{J_{1}}{2}\frac{d\varphi_{b}(t)}{dt}\cos\left(  \frac{\varphi
_{b}(t)-\varphi_{b}(t_{0})}{2}\right)  \sin\left(  \gamma_{b}(t,t_{0})\right)
\nonumber\\
&  -J_{1}\Vert b(t)\Vert\Omega_{b}(t)\sin\left(  \frac{\varphi_{b}%
(t)-\varphi_{b}(t_{0})}{2}\right)  \cos\left(  \gamma_{b}(t,t_{0})\right)
\nonumber\\
&  =-\left(  J_{1}\Vert b(t)\Vert\Omega_{b}(t)+\frac{1}{2}\frac{d\varphi
_{b}(t)}{dt}\right)  \sin\left(  \frac{\varphi_{b}(t)-\varphi_{b}(t_{0})}%
{2}\right)  \cos\left(  \gamma_{b}(t,t_{0})\right) \nonumber\\
&  -\left(  \Vert b(t)\Vert\Omega_{b}(t)+\frac{J_{1}}{2}\frac{d\varphi_{b}%
(t)}{dt}\right)  \cos\left(  \frac{\varphi_{b}(t)-\varphi_{b}(t_{0})}%
{2}\right)  \sin\left(  \gamma_{b}(t,t_{0})\right)  .\nonumber
\end{align}
Similarly, for the derivatives of $\widetilde{u}(t)=(\widetilde{u}%
_{1}(t),\widetilde{u}_{2}(t),\widetilde{u}_{3}(t))\in\mathbb{R}^{3},$ given by
(\ref{70}), we find
\begin{align}
\frac{d}{dt}\widetilde{u}_{1}(t)  &  =-J_{2}\Vert b(t)\Vert\Omega_{b}%
(t)\cos\left(  \frac{\varphi_{b}(t)+\varphi_{b}(t_{0})}{2}\right)  \cos
\gamma_{b}(t,t_{0})\tag{A2}\\
&  +\frac{J_{2}}{2}\frac{d\varphi_{b}(t)}{dt}\sin\left(  \frac{\varphi
(t)+\varphi(t_{0})}{2}\right)  \sin\gamma_{b}(t,t_{0}),\nonumber\\
\frac{d}{dt}\widetilde{u}_{2}(t)  &  =-J_{2}\Vert b(t)\Vert\Omega_{b}%
(t)\sin\left(  \frac{\varphi_{b}(t)+\varphi_{b}(t_{0})}{2}\right)  \cos
\gamma_{b}(t,t_{0})\nonumber\\
&  -\frac{J_{2}}{2}\frac{d\varphi_{b}(t)}{dt}\cos\left(  \frac{\varphi
(t)+\varphi(t_{0})}{2}\right)  \sin\gamma_{b}(t,t_{0}),\nonumber\\
\frac{d}{dt}\widetilde{u}_{3}(t)  &  =-\left(  J_{1}\Vert b(t)\Vert\Omega
_{b}(t)+\frac{1}{2}\frac{d\varphi_{b}(t)}{dt}\right)  \cos\left(
\frac{\varphi_{b}(t)-\varphi_{b}(t_{0})}{2}\right)  \cos\gamma_{b}%
(t,t_{0})\nonumber\\
&  +\left(  \Vert b(t)\Vert\Omega_{b}(t)+\frac{J_{1}}{2}\frac{d\varphi_{b}%
(t)}{dt}\right)  \sin\left(  \frac{\varphi_{b}(t)-\varphi_{b}(t_{0})}%
{2}\right)  \sin\gamma_{b}(t,t_{0}).\nonumber
\end{align}

Next: (i) substituting (\ref{70}) into the terms standing on the right-hand
sides of the equations in (\ref{51}); (ii) expressing $b_{1}(t),$ $b_{2}(t),$
$b_{3}(t)$ in spherical coordinates; (iii) using the trigonometric addition
theorems and the explicit expressions for $J_{1}$, $J_{2}$ and $\Omega_{b}(t)$
(see (\ref{66}) and (\ref{67})), we derive the following expressions:

\begin{itemize}
\item for the right-hand side of the first differential equation in (\ref{51})%
\begin{align}
&  b_{1}(t)\widetilde{u}_{1}(t)+b_{2}(t)\widetilde{u}_{2}(t)+b_{3}%
(t)\widetilde{u}_{3}(t)\tag{A3}\\
&  =-\left(  J_{1}\Vert b(t)\Vert\Omega_{b}(t)+\frac{1}{2}\frac{d\varphi
_{b}(t)}{dt}\right)  \sin\left(  \frac{\varphi_{b}(t)-\varphi_{b}(t_{0})}%
{2}\right)  \cos\left(  \gamma_{b}(t,t_{0})\right) \nonumber\\
&  -\left(  \Vert b(t)\Vert\Omega_{b}(t)+\frac{J_{1}}{2}\frac{d\varphi_{b}%
(t)}{dt}\right)  \cos\left(  \frac{\varphi_{b}(t)-\varphi_{b}(t_{0})}%
{2}\right)  \sin\left(  \gamma_{b}(t,t_{0})\right)  ;\nonumber
\end{align}

\item for the right-hand side of the second differential equation in
(\ref{51})
\begin{align}
&  -b_{1}(t)u_{0}(t)+b_{2}(t)\widetilde{u}_{3}(t)-b_{3}(t)\widetilde{u}%
_{2}(t)\tag{A4}\\
&  =-J_{2}\Vert b(t)\Vert\Omega_{b}(t)\cos\left(  \frac{\varphi_{b}%
(t)+\varphi_{b}(t_{0})}{2}\right)  \cos\left(  \gamma_{b}(t,t_{0})\right)
\nonumber\\
&  +\frac{J_{2}}{2}\frac{d\varphi_{b}(t)}{dt}\sin\left(  \frac{\varphi
(t)+\varphi(t_{0})}{2}\right)  \sin\left(  \gamma_{b}(t,t_{0})\right)
;\nonumber
\end{align}

\item for the right-hand sides of the third and the fourth differential
equations in (\ref{51}):
\begin{align}
&  -b_{2}(t)u_{0}(t)+b_{3}(t)\widetilde{u}_{1}(t)-b_{1}(t)\widetilde{u}%
_{3}(t)\tag{A5}\\
&  =-J_{2}\Vert b(t)\Vert\Omega_{b}(t)\sin\left(  \frac{\varphi_{b}%
(t)+\varphi_{b}(t_{0})}{2}\right)  \cos\left(  \gamma_{b}(t,t_{0})\right)
\nonumber\\
&  -\frac{J_{2}}{2}\frac{d\varphi_{b}(t)}{dt}\cos\left(  \frac{\varphi
(t)+\varphi(t_{0})}{2}\right)  \sin\left(  \gamma_{b}(t,t_{0})\right)
\nonumber
\end{align}
and
\begin{align}
&  -b_{3}(t)u_{0}(t)+b_{1}(t)\widetilde{u}_{2}(t)-b_{2}(t)\widetilde{u}%
_{1}(t)\tag{A6}\\
&  =-\left(  J_{1}\Vert b(t)\Vert\Omega_{b}(t)+\frac{1}{2}\frac{d\varphi
_{b}(t)}{dt}\right)  \cos\left(  \frac{\varphi_{b}(t)-\varphi_{b}(t_{0})}%
{2}\right)  \cos\left(  \gamma_{b}(t,t_{0})\right) \nonumber\\
&  +\left(  \Vert b(t)\Vert\Omega_{b}(t)+\frac{J_{1}}{2}\frac{d\varphi_{b}%
(t)}{dt}\right)  \sin\left(  \frac{\varphi_{b}(t)-\varphi_{b}(t_{0})}%
{2}\right)  \sin\left(  \gamma_{b}(t,t_{0})\right)  .\nonumber
\end{align}

\end{itemize}

Clearly, the expressions for $\frac{d}{dt}u_{0}(t),$ $\frac{d}{dt}%
\widetilde{u}_{1}(t),$ $\frac{d}{dt}\widetilde{u}_{2}(t),$ $\frac{d}%
{dt}\widetilde{u}_{3}(t)$, derived in (A1) and (A2), coincide with the
corresponding expressions in (A3)--(A6). This proves that functions (\ref{70})
constitute the solutions to the Cauchy problem (\ref{51}), equivalently,
(\ref{50}).

Taking into account that (see in Section 2) the unitary evolution operator
$U_{H}(t,s)=u_{0}(t,s)\mathbb{I+}i\widetilde{u}(t,s)\cdot\sigma,$ for each
$s\in\lbrack t,t_{0}],$ let us now prove that solutions (\ref{70}) satisfy the
cocycle property (\ref{2.1}) for $U_{H}(t,t_{0}).$ In terms of $u_{0}(t,s),$
$\widetilde{u}(t,s),$ the cocycle property leads to relations (\ref{61.1}),
which read:%

\begin{align}
u_{0}(t,s)u_{0}(s,t_{0})-\widetilde{u}(t,s)\cdot\widetilde{u}(s,t_{0})  &
=u_{0}(t,t_{0}),\tag{A7}\\
u_{0}(t,s)\widetilde{u}(s,t_{0})+u_{0}(s,t_{0})\widetilde{u}%
(t,s)-\widetilde{u}(t,s)\times\widetilde{u}(s,t_{0})  &  =\widetilde{u}%
(t,t_{0}).\nonumber
\end{align}

Substituting solutions (\ref{70}) into (A7), applying the addition rules for
trigonometric functions and taking into account that $J_{1}^{2}+J_{2}^{2}=1$,
for the left-hand side of the first equation in (A7), we derive:
\begin{align}
&  u_{0}(t,s)u_{0}(s,t_{0})-\widetilde{u}(t,s)\cdot\widetilde{u}%
(s,t_{0})\tag{A8}\\
&  =\cos\left(  \gamma_{b}(t,s)\right)  \cos\left(  \gamma_{b}(s,t_{0}%
)\right)  \cos\left(  \frac{\varphi_{b}(t)-\varphi_{b}(t_{0})}{2}\right)
\nonumber\\
&  -J_{1}\sin\left(  \frac{\varphi_{b}(t)-\varphi_{b}(t_{0})}{2}\right)
{\LARGE [}\cos\left(  \gamma_{b}(t,s)\right)  \sin\left(  \gamma_{b}%
(s,t_{0})\right) \nonumber\\
&  +\sin\left(  \gamma_{b}(t,s)\right)  \cos\left(  \gamma_{b}(s,t_{0}%
)\right)  {\LARGE ]}\nonumber\\
&  -\sin\left(  \gamma_{b}(t,s)\right)  \sin\left(  \gamma_{b}(s,t_{0}%
)\right)  \cos\left(  \frac{\varphi_{b}(t)-\varphi_{b}(t_{0})}{2}\right)
\nonumber\\
&  =\cos\left(  \frac{\varphi_{b}(t)-\varphi_{b}(t_{0})}{2}\right)
\cos\left(  \gamma_{b}(t,t_{0})\right) \nonumber\\
&  -J_{1}\sin\left(  \frac{\varphi_{b}(t)-\varphi_{b}(t_{0})}{2}\right)
\sin\left(  \gamma_{b}(t,t_{0})\right)  . \tag{A9}%
\end{align}
By the same procedure, for the left-hand sides of the remaining equations in
(A7), we have
\begin{align}
&  u_{0}(t,s)\widetilde{u}_{1}(s,t_{0})+u_{0}(s,t_{0})\widetilde{u}%
_{1}(t,s)-\left(  \widetilde{u}(t,s)\times\widetilde{u}(s,t_{0})\right)
_{1}\tag{A10}\\
&  =-J_{2}{\LARGE [}\sin\left(  \gamma_{b}(t,s)\right)  \cos\left(  \gamma
_{b}(s,t_{0})\right) \nonumber\\
&  +\cos\left(  \gamma_{b}(t,s)\right)  \sin\left(  \gamma_{b}(s,t_{0}%
)\right)  {\LARGE ]}\cos\left(  \frac{\varphi_{b}(t)+\varphi_{b}(t_{0})}%
{2}\right) \nonumber\\
&  +J_{1}J_{2}\sin\left(  \gamma_{b}(t,s)\right)  \sin\left(  \gamma
_{b}(s,t_{0})\right)  {\LARGE [}\sin\left(  \frac{\varphi_{b}(t)-\varphi
_{b}(s)}{2}\right)  \cos\left(  \frac{\varphi_{b}(s)+\varphi_{b}(t_{0})}%
{2}\right) \nonumber\\
&  +\cos\left(  \frac{\varphi_{b}(t)+\varphi_{b}(s)}{2}\right)  \cos\left(
\frac{\varphi_{b}(s)-\varphi_{b}(t_{0})}{2}\right) \nonumber\\
&  +\cos\left(  \frac{\varphi_{b}(t)-\varphi_{b}(s)}{2}\right)  \sin\left(
\frac{\varphi_{b}(s)+\varphi_{b}(t_{0})}{2}\right) \nonumber\\
&  -\sin\left(  \frac{\varphi_{b}(t)+\varphi_{b}(s)}{2}\right)  \cos\left(
\frac{\varphi_{b}(s)-\varphi_{b}(t_{0})}{2}\right)  {\LARGE ]}\nonumber\\
&  =-J_{2}\cos\left(  \frac{\varphi_{b}(t)+\varphi_{b}(t_{0})}{2}\right)
\sin\left(  \gamma_{b}(t,t_{0})\right)  \tag{A11}%
\end{align}
and
\begin{align}
&  u_{0}(t,s)\widetilde{u}_{2}(s,t_{0})+u_{0}(s,t_{0})\widetilde{u}%
_{2}(t,s)-\left(  \widetilde{u}(t,s)\times\widetilde{u}(s,t_{0})\right)
_{2}\tag{A12}\\
&  =-J_{2}[\sin\left(  \gamma_{b}(t,s)\right)  \cos\left(  \gamma_{b}%
(s,t_{0})\right) \nonumber\\
&  +\cos\left(  \gamma_{b}(t,s)\right)  \sin\left(  \gamma_{b}(s,t_{0}%
)\right)  ]\sin\left(  \frac{\varphi_{b}(t)+\varphi_{b}(t_{0})}{2}\right)
\nonumber\\
&  +J_{1}J_{2}\sin\left(  \gamma_{b}(t,s)\right)  \sin\left(  \gamma
_{b}(s,t_{0})\right)  {\LARGE [}\sin\left(  \frac{\varphi_{b}(t)-\varphi
_{b}(s)}{2}\right)  \sin\left(  \frac{\varphi_{b}(s)+\varphi_{b}(t_{0})}%
{2}\right) \nonumber\\
&  +\sin\left(  \frac{\varphi_{b}(t)+\varphi_{b}(s)}{2}\right)  \sin\left(
\frac{\varphi_{b}(s)-\varphi_{b}(t_{0})}{2}\right) \nonumber\\
&  -\cos\left(  \frac{\varphi_{b}(t)-\varphi_{b}(s)}{2}\right)  \cos\left(
\frac{\varphi_{b}(s)+\varphi_{b}(t_{0})}{2}\right) \nonumber\\
&  +\cos\left(  \frac{\varphi_{b}(t)+\varphi_{b}(s)}{2}\right)  \cos\left(
\frac{\varphi_{b}(s)-\varphi_{b}(t_{0})}{2}\right)  {\LARGE ]}\nonumber\\
&  =-J_{2}\sin\left(  \frac{\varphi_{b}(t)+\varphi_{b}(t_{0})}{2}\right)
\sin\left(  \gamma_{b}(t,t_{0})\right)  , \tag{A13}%
\end{align}
and
\begin{align}
&  u_{0}(t,s)\widetilde{u}_{3}(s,t_{0})+u_{0}(s,t_{0})\widetilde{u}%
_{3}(t,s)-\left(  \widetilde{u}(t,s)\times\widetilde{u}(s,t_{0})\right)
_{3}\tag{A14}\\
&  =-\cos\left(  \gamma_{b}(t,s)\right)  \cos\left(  \gamma_{b}(s,t_{0}%
)\right)  \sin\left(  \frac{\varphi_{b}(t)-\varphi_{b}(t_{0})}{2}\right)
\nonumber\\
&  -J_{1}\cos\left(  \frac{\varphi_{b}(t)-\varphi_{b}(t_{0})}{2}\right)
[\cos\left(  \gamma_{b}(t,s)\right)  \sin\left(  \gamma_{b}(s,t_{0})\right)
\nonumber\\
&  +\sin\left(  \gamma_{b}(t,s)\right)  \cos\left(  \gamma_{b}(s,t_{0}%
)\right)  ]\nonumber\\
&  +J_{1}^{2}\sin\left(  \gamma_{b}(t,s)\right)  \sin\left(  \gamma
_{b}(s,t_{0})\right)  \sin\left(  \frac{\varphi_{b}(t)-\varphi_{b}(t_{0})}%
{2}\right) \nonumber\\
&  +J_{2}^{2}\sin\left(  \gamma_{b}(t,s)\right)  \sin\left(  \gamma
_{b}(s,t_{0})\right)  \sin\left(  \frac{\varphi_{b}(t)-\varphi_{b}(t_{0})}%
{2}\right) \nonumber\\
&  =-J_{1}\cos\left(  \frac{\varphi_{b}(t)-\varphi_{b}(t_{0})}{2}\right)
\sin\left(  \gamma_{b}(t,t_{0})\right) \nonumber\\
&  -\sin\left(  \frac{\varphi_{b}(t)-\varphi_{b}(t_{0})}{2}\right)
\cos\left(  \gamma_{b}(t,t_{0})\right)  . \tag{A15}%
\end{align}
The comparison of (A9), (A11), (A13) and (A15) with the expressions for
functions $u_{0}(t),\widetilde{u}(t)$ in Theorem 3 proves that the unitary
evolution qubit operator $U_{H}(t,t_{0}),$ specified in Theorem 3, satisfies
the cocycle property (\ref{61.1}).

This concludes the proof of Theorem 3.

\section{Appendix B}

In this section, we show that the gradient of $K_{3}(r)$ is given by
(\ref{47.1}). By (\ref{47}) we have
\begin{align}
\frac{\partial K_{3}(r)}{\partial r}  &  =-i\sqrt{2}\sum\limits_{k=0,1,2}%
{\LARGE [}\frac{r}{\Vert r\Vert_{\mathbb{R}^{8}}}\sin\left(  \phi(r)+2\pi
k/3\right) \tag{B1}\\
&  +\Vert r\Vert_{\mathbb{R}^{8}}\frac{\partial\phi(r)}{\partial r}\cos\left(
\phi(r)+2\pi k/3\right)  {\LARGE ]}\nonumber\\
&  \times\exp\left\{  -i\sqrt{2}\Vert r\Vert_{\mathbb{R}^{8}}\sin\left(
\phi(r)+2\pi k/3\right)  \right\}  .\nonumber
\end{align}
Using further (\ref{46.1}), we derive
\begin{equation}
\frac{\partial\phi(r)}{\partial r}=-\frac{1}{\cos(3\phi(r))}\left(  \frac
{r}{\Vert r\Vert_{\mathbb{R}^{8}}^{2}}\sin(3\phi(r))+\frac{\sqrt{3}}{2}%
\frac{1}{\Vert r\Vert_{\mathbb{R}^{8}}^{3}}\frac{\partial}{\partial r}\left(
\text{det}(r\cdot\Lambda)\right)  \right)  , \tag{B2}%
\end{equation}
where
\begin{align}
\text{det}(r\cdot\Lambda)  &  =2(r^{(1)}r^{(4)}r^{(6)}+r^{(1)}r^{(5)}%
r^{(7)}+r^{(2)}r^{(5)}r^{(6)}-r^{(2)}r^{(4)}r^{(7)})\tag{B3}\\
&  +\frac{1}{\sqrt{3}}r^{(8)}\left(  2(r^{(1)})^{2}+2(r^{(2)})^{2}%
+2(r^{(3)})^{2}-(r^{(4)})^{2}-(r^{(5)})^{2}-(r^{(6)})^{2}-(r^{(7)})^{2}\right)
\nonumber\\
&  +r^{(3)}\left(  (r^{(4)})^{2}+(r^{(5)})^{2}-(r^{(6)})^{2}-(r^{(7)}%
)^{2}\right)  -\frac{2}{3\sqrt{3}}(r^{(8)})^{3}.\nonumber
\end{align}
Taking into account that the symmetric structure constants $d_{ijk}$ of
$\text{SU}(3)$ have the form (see, e.g., in \cite{f}):%
\begin{align}
d_{146}  &  =d_{157}=d_{256}=d_{344}=d_{355}=-d_{247}=-d_{366}=-d_{377}%
=\frac{1}{2},\tag{B4}\\
d_{118}  &  =d_{228}=d_{338}=-d_{888}=-2d_{448}=-2d_{558}=-2d_{668}%
=-2d_{778}=\frac{1}{\sqrt{3}},\nonumber
\end{align}
we derive%
\begin{subequations}
\begin{equation}
\frac{\partial}{\partial r^{(l)}}\left(  \text{det}(r\cdot\Lambda)\right)
=2\sum\limits_{i,j=1}^{8}r^{(i)}r^{(j)}d_{ijl}. \tag{B5}%
\end{equation}
Hence, (B2) reduces to
\end{subequations}
\begin{equation}
\frac{\partial\phi(r)}{\partial r}=-\frac{1}{\cos(3\phi(r))}\left(  \frac
{r}{\Vert r\Vert_{\mathbb{R}^{8}}^{2}}\sin(3\phi(r))+\sqrt{3}\frac{p(r)}{\Vert
r\Vert_{\mathbb{R}^{8}}}\right)  \tag{B6}%
\end{equation}
and, for (B1), we obtain
\begin{align}
\frac{\partial K_{3}(r)}{\partial r}  &  =-i\sqrt{2}\sum\limits_{k=0,1,2}%
{\LARGE [}\frac{r}{\Vert r\Vert_{\mathbb{R}^{8}}}\left(  \sin(\phi(r)+2\pi
k/3)-\sin(3\phi(r))\frac{\cos(\phi(r)+2\pi k/3)}{\cos(3\phi(r))}\right)
\tag{B7}\\
&  -\sqrt{3}p(r)\frac{\cos(\phi(r)+2\pi k/3)}{\cos(3\phi(r))}{\LARGE ]}%
\times\exp\left\{  -i\sqrt{2}\Vert r\Vert_{\mathbb{R}^{8}}\sin(\phi(r)+2\pi
k/3)\right\}  .\nonumber
\end{align}
Noting that, on the right-hand side of (B7), $\cos\left(  3\phi(r)\right)
=\cos\left(  3\left(  \phi(r)+2\pi k/3\right)  \right)  $,%
\begin{align}
-  &  \frac{\cos(\phi(r)+2\pi k/3)}{\cos(3(\phi(r)+2\pi k/3))}=-\frac{1}%
{4\cos^{2}(\phi(r)+2\pi k/3)-3}\tag{B8}\\
&  =\frac{1}{1-2\cos(2(\phi(r)+2\pi k/3))}\nonumber
\end{align}
and
\begin{align}
&  \sin\left(  \phi(r)+2\pi k/3\right)  -\frac{\sin(3\phi(r))}{\cos(3\phi
(r))}\cos\left(  \phi(r)+2\pi k/3\right) \tag{B10}\\
&  =-\frac{\sin\left(  2(\phi(r)+2\pi k/3)\right)  }{\cos(3\phi(r))}%
=\frac{2\sin(\left(  \phi(r)+2\pi k/3\right)  }{1-2\cos(\left(  2(\phi(r)+2\pi
k/3)\right)  },\nonumber
\end{align}
for the second and the first terms in the right-hand side of (B7), we come
correspondingly to the following expressions%
\begin{align}
&  -i\sqrt{6}p(r)\sum\limits_{k=0,1,2}\frac{\exp\left\{  -i\sqrt{2}\Vert
r\Vert_{\mathbb{R}^{8}}\sin(\phi(r)+2\pi k/3)\right\}  }{1-2\cos
(2(\phi(r)+2\pi k/3))}\tag{B11}\\
&  =-i\sqrt{6}F_{1}(r)p(r)\nonumber
\end{align}
and
\begin{align}
&  -i\sqrt{2}\frac{r}{\Vert r\Vert_{\mathbb{R}^{8}}}\sum\limits_{k=0,1,2}%
\sin\left(  \phi(r)+2\pi k/3\right)  \frac{\exp\left\{  -i\sqrt{2}\left\Vert
r\right\Vert _{\mathbb{R}^{8}}\sin(\phi(r)+2\pi k/3)\right\}  }{1-2\cos
(\left(  2(\phi(r)+2\pi k/3)\right)  }\tag{B12}\\
&  =-i\sqrt{6}F_{2}(r)\frac{r}{\Vert r\Vert_{\mathbb{R}^{8}}},\nonumber
\end{align}
where functions $F_{1}(r)$ and $F_{2}(r)$ are given by (\ref{47.2}).

Relations (B7), (B11) and (B12) prove Eq. (\ref{47.1}).

\end{document}